# Preservation of photon statistics upon selective molecular modulation in hollow-core fibers

Arturo Mena[1*], Pau Arcos[1], María Sánchez-Hernández[1], Benoit Debord[2], Frédéric Gérôme[2], Fetah Benabid[2], Nicolas Y. Joly[3,4], Oskar Arrizabalaga[5], Joseba Zubia[1,6], David Novoa[1,6,7*]

[1]Department of Communications Engineering, University of the Basque Country (EHU), Bilbao, 48013, Spain.
[2]GPPMM Group, XLIM Institute, CNRS UMR 7252, University of Limoges, Limoges, 87100, France.
[3]Department of Physics, University of Erlangen-Nürnberg, Erlangen, 91058, Germany.
[4]Max Planck Institute for the Science of Light, Erlangen, 91058, Germany.
[5]Department of Applied Mathematics, University of the Basque Country (EHU), Bilbao, 48013, Spain.
[6]EHU Quantum Center, University of the Basque Country (EHU), Bilbao, Spain.
[7]IKERBASQUE, Basque Foundation for Science, Bilbao, 48009, Spain.

**Abstract**

Preserving the underlying statistical properties of light during frequency transduction is essential for interfacing spectrally disparate quantum systems. To do this successfully, the process needs to be highly efficient, broadband and noiseless. Despite significant progress in this field, satisfying all these conditions remains challenging. Recently, molecular modulation-based transduction in hollow-core fibers has emerged as a promising approach, where biphoton correlations were proven to be preserved during frequency up-conversion. However, beyond this previous demonstration of correlation-maintaining conversion, whether molecular modulation can preserve more general statistical properties of light remains experimentally unexplored. Here, we demonstrate the conservation of photon-number statistics during selective frequency transduction by Raman molecular modulation in a hydrogen-filled hollow-core fiber. As a proof of concept, weak 585-nm signals are selectively converted to 773 nm in the near-infrared and to 471 nm in the blue spectral region with near-unity efficiencies. Phase matching between the generated coherence waves and the transitions of interest is achieved through the effective dispersion of the system, enabling independent selection of the frequency up- and down-conversion processes. The preservation of the photon-number statistics of the shifted signals, along with the large frequency shifts, broad tunability and compatibility with quantum-optical fields, provides a versatile fiber-based platform for quantum frequency transduction. In particular, the demonstrated interfacing between visible and near-infrared wavelengths using low pump energies and low gas pressures offers a route towards connecting spectrally distinct quantum systems, including quantum emitters, sensors and atomic quantum memories, within hybrid quantum networks.

## 1 INTRODUCTION

Quantum technologies are rapidly progressing from demonstrations of individual quantum devices towards increasingly complex systems in which distinct quantum platforms must operate together[1, 2]. The resulting diversity of light-based platforms, including quantum emitters, memories, processors and sensors[2, 3], is accompanied by a variety of operating wavelengths. Connecting such heterogeneous devices is therefore a central challenge for the development of distributed quantum technologies and, ultimately, a quantum internet[4]. Photons are natural carriers of quantum information between remote nodes. Consequently, efficient interconnection requires frequency conversion across widely separated spectral domains. Quantum frequency transduction has therefore emerged as an important enabling technology for hybrid quantum networks, with demonstrations spanning nonlinear optical processes in bulk crystals[5] and specialty solid-core fibers[6], optomechanical interactions[7], and other platforms[8]. Beyond conversion efficiency, however, a useful quantum transducer must introduce minimal noise and maintain the statistical properties of the input quantum state[9].

In recent years, hollow-core optical fibers (HCFs) have demonstrated to be a promising vehicle for lightwave quantum technologies[10-14]. In the context of quantum networking, their unique characteristics such as high damage threshold[15] and ultralow-loss broadband guidance[16-18], have facilitated the development of quantum memories[19, 20] and the coexistence of classical and quantum channels in quantum communications[21, 22]. Additionally, when selectively filled with gases, HCFs have also enabled access to a range of nonlinear optical processes spanning wide spectral regions[23-28], which gave rise, e.g., to the development of quantum light sources[29-31] and frequency transducers[32, 33].

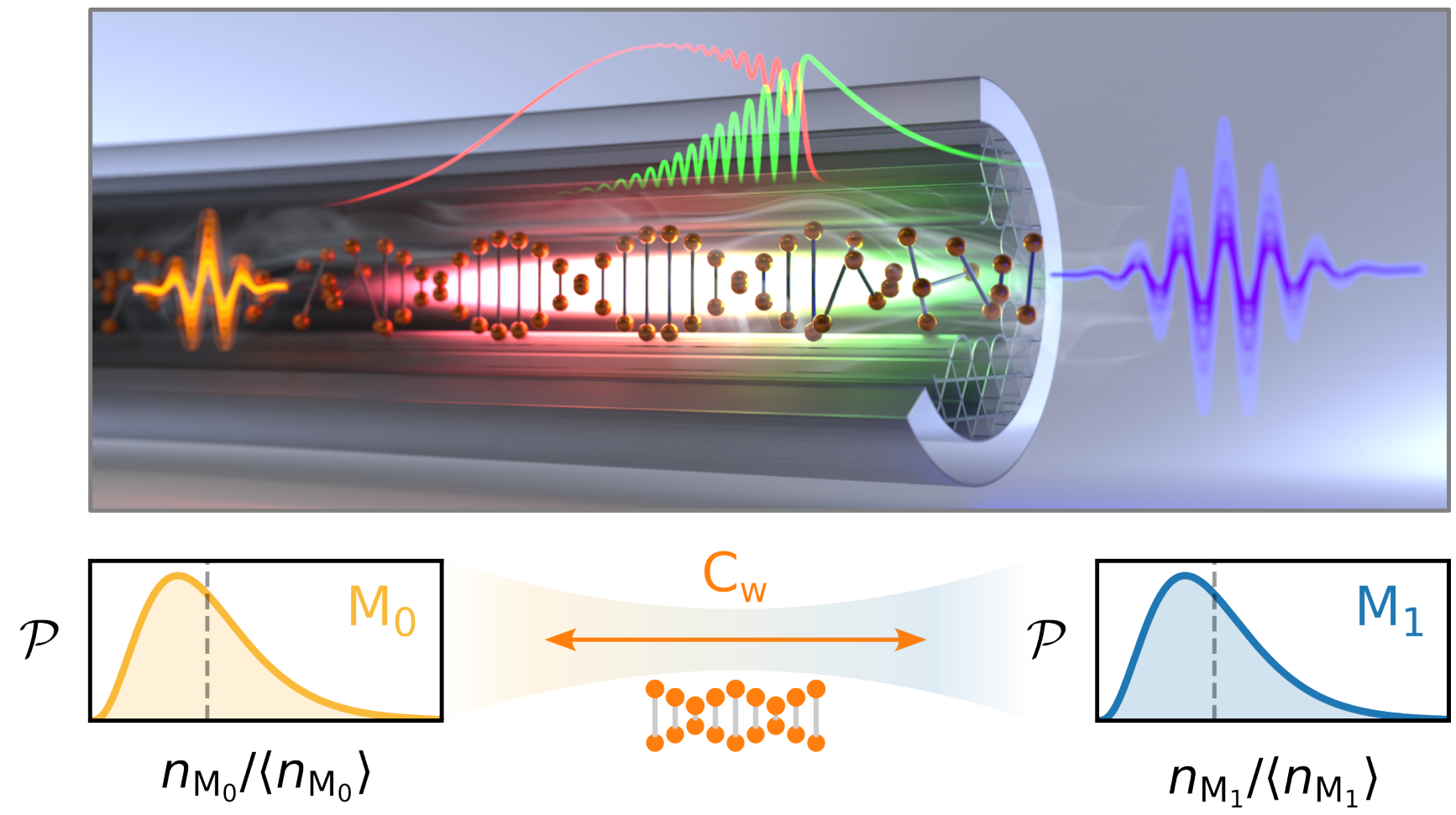

**Fig. 1** Artistic depiction of frequency transduction by molecular modulation in a hydrogen-filled HCF. The pump pulse in green drives a vibrational transition of the hydrogen gas. Together with its frequency down-shifted Stokes component depicted in red, they drive a wave of synchronized molecular vibrations: the coherence wave, $C_w$. Then, a phase-matched mixing pulse $M_0$ is inelastically scattered by this molecular coherence to the converted frequency pulse $M_1$, which inherits the photon number statistics $\mathcal{P}$ of the original pulse when the process is highly efficient.

In this regard, Raman scattering-based molecular modulation offers a particularly interesting approach to the problem of highly efficient frequency transduction of low-intensity light[32, 34]. In this process, a collective excitation of synchronized molecular vibrations known as coherence wave is prepared by driving an appropriate Raman transition[35, 36]. A weak optical field is subsequently coupled to this coherence wave, generating frequency-shifted sidebands through a parametric interaction (see Fig. 1). Unlike conventional Raman scattering, frequency conversion using molecular modulation can be operated below the Raman threshold because the molecular coherence wave has been previously prepared via stimulated Raman scattering (SRS), allowing weak optical fields even down to the single-photon level to be selectively translated in frequency with near-unit efficiency[32, 37].

An important question nevertheless remains: to what extent does molecular modulation preserve the quantum statistics of the light being converted? This approach has previously been shown to preserve biphoton second-order correlations[32, 38]. However, quantum networks and photonic quantum technologies also require the faithful transmission and processing of more general states, for which the full photon-number statistics can be essential[39, 40]. Establishing whether selective molecular modulation transfers these statistics is therefore crucial for assessing its suitability as a viable quantum transduction mechanism.

Here, we investigate the preservation of photon statistics upon selective frequency up- and down-conversion in gas-filled HCFs. We use molecular modulation to translate weak optical pulses between distinct spectral regions, demonstrating frequency transduction while operating below the Raman threshold. We characterize the photon statistics of the original and frequency-shifted light and show that the highly-efficient modulation process can preserve these statistics despite the large spectral translation. Particularly, as proof of concept, we focus on the transduction of 585 nm visible wavelengths with low intensities into the near infrared at ~773 nm by using a single shift of ~125 THz provided by the Q(1) vibrational transition of molecular hydrogen. Similarly, using the same system we also demonstrate up-conversion to blue wavelengths (~471 nm), which are of interest for nitrogen-vacancy center technology[41]. Together with the ability of molecular modulation to convert broadband and single-photon light, this provides efficient bridges between quantum technologies working in the visible such as quantum-dot single-photon sources used for sensing and near infrared technologies like rubidium-based quantum memories[42]. Our results establish molecular modulation in HCFs as a fiber-compatible route for frequency-selective manipulation of quantum light, providing a viable alternative for connecting spectrally disparate quantum systems that could be integrated into future hybrid quantum networks.

## 2 RESULTS

### Transduction characterization

As shown in Fig. 2a, the experiment consists of three main stages: generation, transduction and detection (see Materials and Methods for details). The mixing source is built by pressurizing a tubular HCF[43] with nitrogen gas and pumping with 515 nm pulses. As a result, this stage produces yellow Stokes radiation with a wavelength of ~585 nm via vibrational SRS, which is used as mixing light with tunable photon-number statistical properties for characterizing the transduction. Although these statistics follow in general a Bose-Einstein distribution, for the cases where the integration time $T$ of the photodetection system is longer than the coherence time $T_c$ of the mixing pulses, the retrieved photon statistics $\mathcal{P}_\mu$ are best described by the gamma distribution[44]

$$\mathcal{P}_\mu(n) = \frac{\mu^\mu}{\langle n \rangle \Gamma(\mu)} \left(\frac{n}{\langle n \rangle}\right)^{\mu-1} \exp\left(-\mu \frac{n}{\langle n \rangle}\right), \quad (1)$$

with Γ denoting the gamma function. This distribution, with an average photon number $\langle n \rangle$, considers that there are $\mu \sim T/T_c$ statistically independent modes being detected and has been experimentally observed by previous works on Raman scattering dynamics[45]. Therefore, by simply changing the pump power we can access states with different statistics as can be seen in Figures 2b and c. The generated power of the 585 nm 1st Stokes increases exponentially with the pump power $P_P^{(N_2)}$, transitioning from positively skewed gamma statistics with larger variance to symmetrical distributions. However, we may note that this exponential behavior begins to saturate when the power of the 2nd Stokes band at ~678 nm becomes relevant and the 1st Stokes is depleted. The resulting 585 nm pulses are attenuated using neutral density filters to powers well below the Raman threshold of the transducer fiber in the second stage of the experiment, so as to ensure that they cannot drive their own molecular coherence. After this, they are combined with the synchronized pump pulses of the next stage at a dichroic mirror for injection into the transducer.

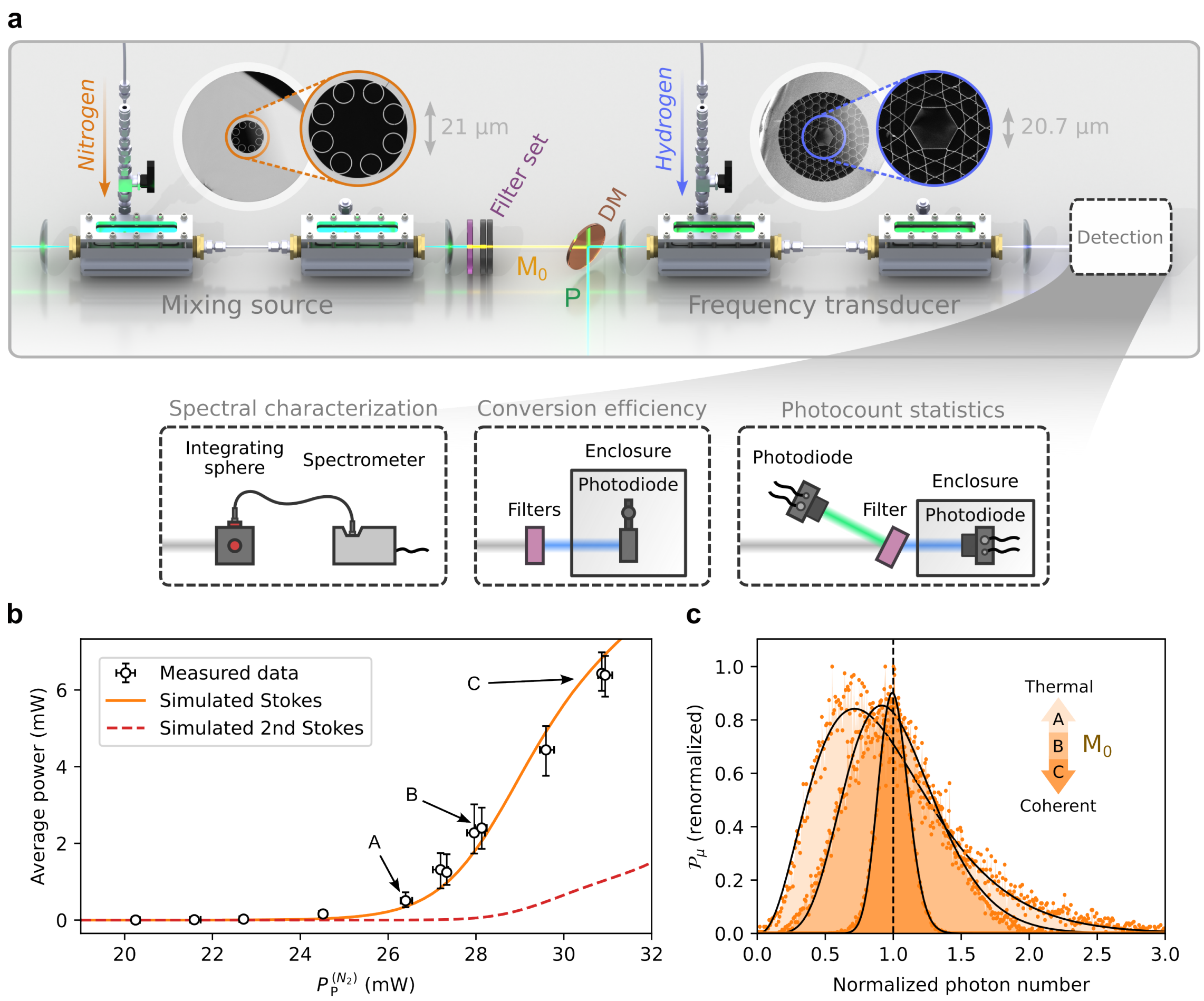

**Fig. 2** Description of the experimental setup. **a** Schematic experimental layout consisting of a mixing state source, the frequency transducer and the detection system, which is readapted depending on the type of characterization. Here, DM is a dichroic mirror used for combining the mixing ($M_0$) and pump (P) frequencies. Scanning electron microscopy images of the HCF cross-sections considered are also shown. **b** Average $M_0$ power at the output of the mixing source without attenuation filters as a function of the average pump power. **c** Experimental photon number statistics of the $M_0$ light at different regimes of operation, normalized with respect to the average photon number of each case and fitted to a gamma distribution (black solid line). The labels A, B and C connect the measurements of **b** and **c**.

In the transduction stage, an ultrathin-walled kagome HCF[23] is pressurized with hydrogen and also pumped at 515 nm with energies optimizing the transduction. The spectral guidance window of this HCF is broadband enough to guide all the wavelengths involved in the dynamics. To determine the kagome HCF capabilities for selective transduction we analyzed the phase-matching conditions according to the dispersion of the system, which is dependent on the cross-sectional geometry of the fiber and the properties of the filling gas. For this purpose, we consider the semiclassical Maxwell-Bloch formalism of molecular modulation as detailed in Materials and Methods, describing both the generation of molecular coherence by SRS and the subsequent transduction of the low-intensity mixing pulses. In this description, the pump beam generates a down-shifted Stokes wavelength at ~655 nm by noise-initiated SRS. The beat-note given by these two frequencies drives the resulting molecular coherence wave, which will travel along the fiber with the pulses having a propagation constant $\beta_{\mathrm{C_w}} = \beta_{\mathrm{P}} - \beta_{\mathrm{S_1}}$ determined by those of the pump ($\beta_{\mathrm{P}}$) and Stokes ($\beta_{\mathrm{S_1}}$) bands. Therefore, a weak mixing signal, unable to generate molecular coherence, will experience inelastic scattering only if the beat-note with the up- or down-converted frequency matches that of the pre-existing coherence wave. That is, defining the dephasing $\vartheta_{m,m-1} = \left(\beta_{\mathrm{P}} - \beta_{\mathrm{S_1}}\right) - \left(\beta_m - \beta_{m-1}\right)$, the frequency up-shifting transduction is optimized when $\vartheta_{\mathrm{M_1,M_0}} = 0$ and the down-shifting when $\vartheta_{\mathrm{M_0,M_{-1}}} = 0$. Here, we have labeled the down-converted, mixing and up-converted frequencies as $\mathrm{M_{-1}}$, $\mathrm{M_0}$ and $\mathrm{M_1}$. Therefore, we find the phase-matching hydrogen pressurization to be ~16.7 bar for frequency up-shifting and ~28.1 bar for down-shifting.

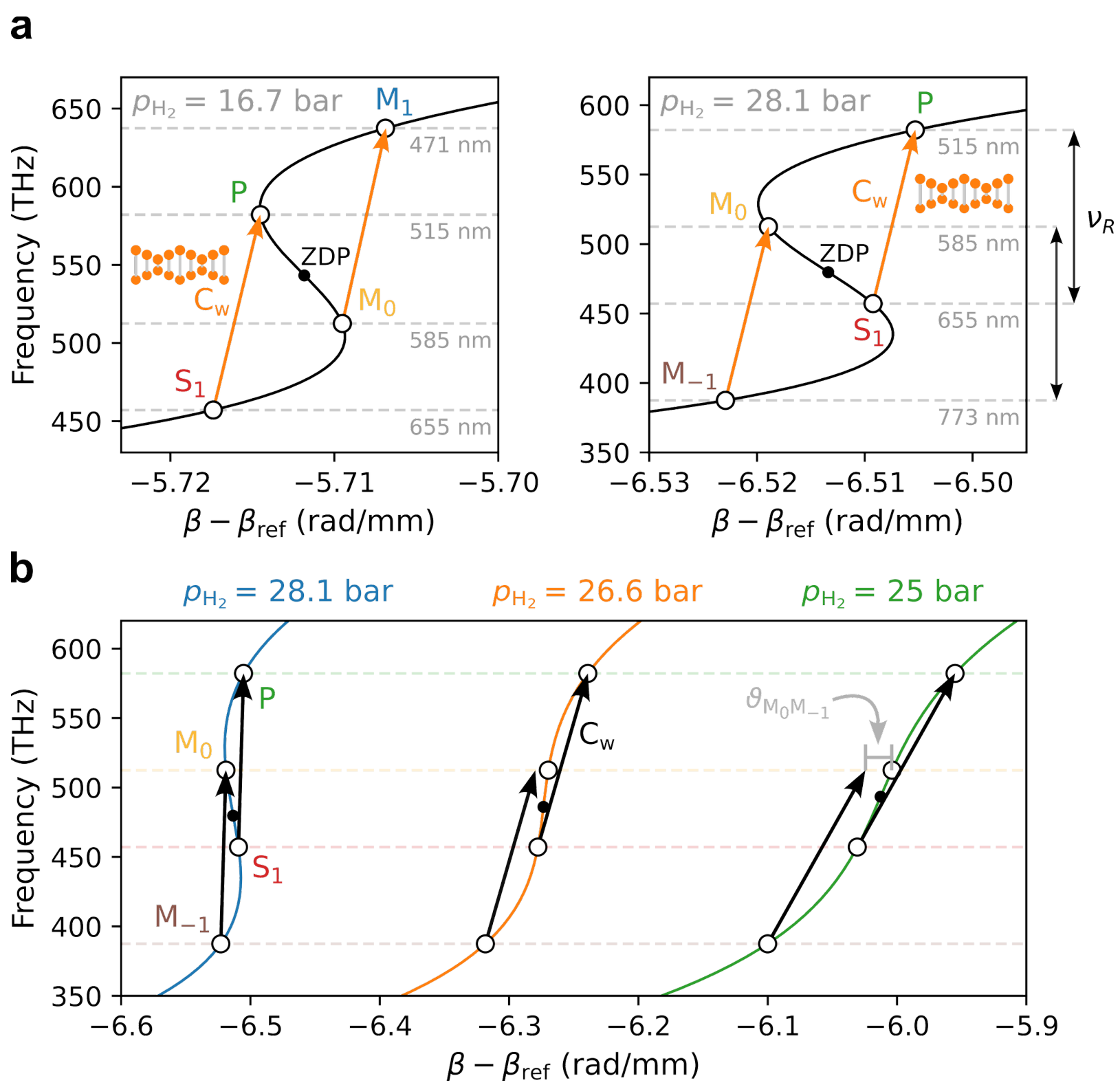

**Fig. 3** Phase-matching dispersion diagrams for frequency up- and down-conversion. **a** Diagrams for the two phase-matching pressures of interest. Here, the coherence wave $\mathrm{C_w}$ driven by the P and $\mathrm{S_1}$ bands is able to connect the mixing and converted frequencies, enabling efficient photon population transfer. The mixing ($\mathrm{M_0}$) converted sidebands are denoted as $\mathrm{M_1}$ for up-conversion and $\mathrm{M_{-1}}$ for down-conversion. The corresponding wavelengths are also indicated below the dashed lines. **b** Down-conversion diagram as the hydrogen pressure $p_{\mathrm{H_2}}$ is decreased and set out of the value of ideal phase-matching. $\vartheta_{\mathrm{M_0,M_{-1}}}$ denotes the dephasing. In both sub-figures, the black dots mark the zero-dispersion points (ZDP) of the system.

The phase-matching conditions can be visualized graphically with aid of dispersion diagrams, presented in Fig. 3 for the up- and down-conversion processes. To magnify the deviation from linear behavior and reveal the underlying S-shaped profile, the propagation constants $\beta$ of the fundamental $LP_{01}$-like mode of the fiber are shifted by the linear reference contribution $\beta_{\mathrm{ref}} = (2\pi\nu/c)n_{\mathrm{eff}}(\nu_{\mathrm{ref}})$, with $\nu$ being the frequency. For better visualization, the arbitrary reference frequency was chosen as $\nu_{\mathrm{ref}} = 1318\ (1180)$ THz in the up-(down-)conversion case. In the diagrams of Fig. 3, the coherence wave is visualized as a vector connecting the P and $S_1$ sidebands. The components of this vector are written as $C_w = \left(\beta_P - \beta_{\mathrm{ref}}(\nu_P) - \beta_{S_1} + \beta_{\mathrm{ref}}(\nu_{S_1}), \nu_R\right)$, where in our case the Raman shift is $\nu_R \sim 125$ THz as mentioned previously. In Fig. 3a, when phase-matching is satisfied, the $C_w$ vector will also connect the $M_0$ and converted sidebands. However, when the pressure is tuned to be different than the phase-matching pressure, as seen in Fig. 3b, the vector will not be able to connect any mixing sideband pair. Note that under phase-matched conditions the transduction can be bidirectional. For example, in up-conversion, the $M_1$ band can play the role of mixing while $M_0$ plays the role of down-converted.

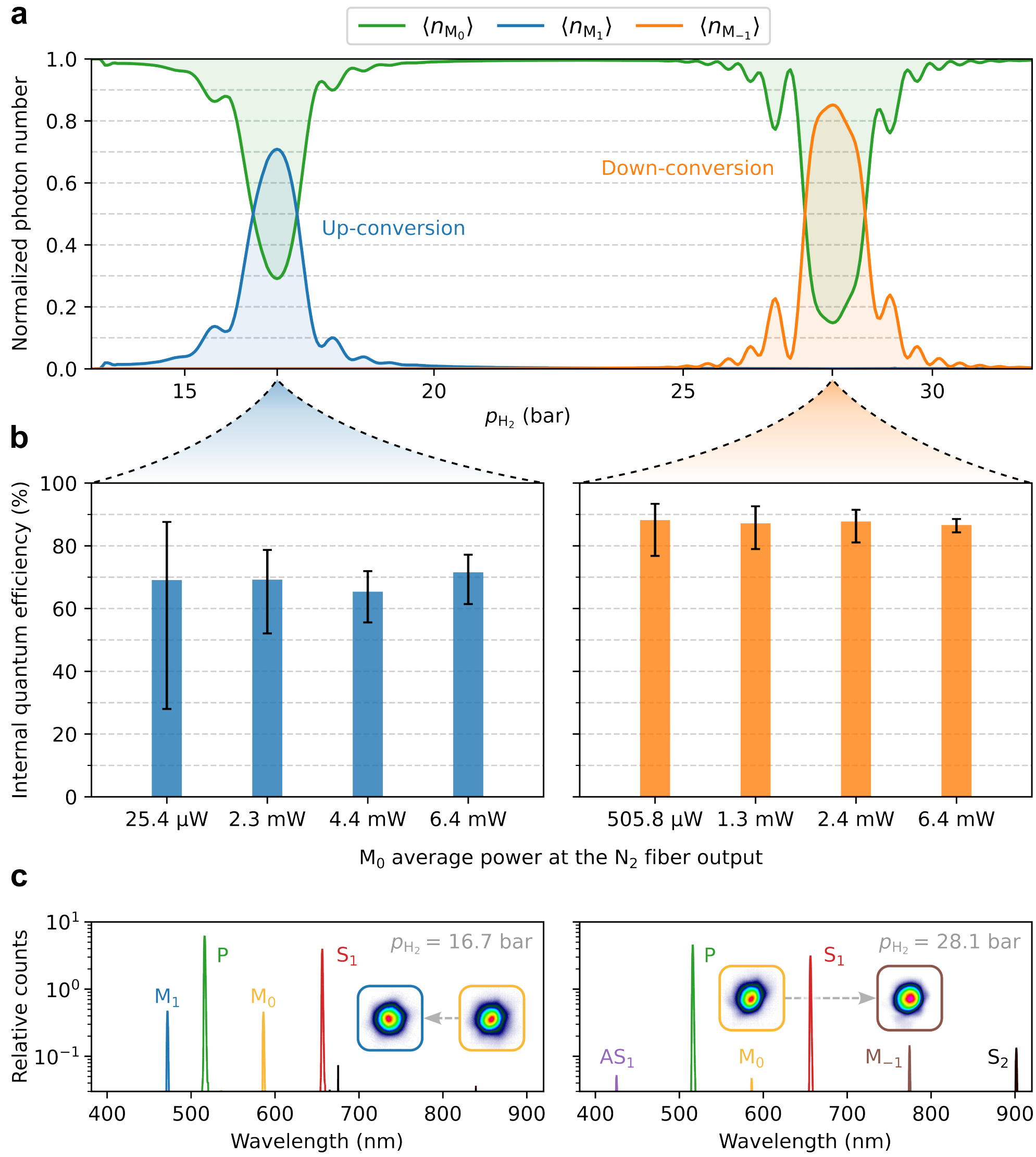

**Fig. 4** Characterization of the frequency transduction system. **a** Maxwell-Bloch simulation of the photon numbers of the mixing sidebands at the output of the transducer fiber during a hydrogen pressure scan. Two clear peaks can be observed at the phase-matching pressures of ~16.7 bar and ~28.1 bar. **b** Experimental measurements of the internal quantum efficiency of the up- (left) and down-conversion (right) processes. **c** Spectral measurements of the wavelengths involved in the frequency transduction. Both the sidebands of the pump and mixing frequency are shown. The mixing sidebands are denoted by $M_m$, with $m = -1, 1$, while the pump (P) sidebands are denoted as $S_l$ for the Stokes bands and $AS_l$ for the anti-Stokes bands, with $l = 1, 2$. Here, the color-coded insets show the near-field mode profiles of the corresponding mixing sidebands.

Fig. 4a showcases a Maxwell-Bloch simulation of the expected photon numbers at the output of the transducer as a function of the hydrogen pressurization. The pump pulse input energy is 0.5 μJ and the presented photon numbers are normalized with respect to the output mixing photon number when the conversion is turned off. Remarkably, the pressure scan exhibits two peaks centered at the pressures predicted by the theory and sufficiently separated so that there is negligible cross-talk (below 40 dB) between the up- and down-conversion processes.

The experimental conversion efficiencies at the predicted optimum pressures for both up- (~16.7 bar) and down-shifting (~28.1 bar) are shown at Fig. 4b. In this figure, the label below each bar denotes the mixing average power at the output of the nitrogen-filled HCF used for its generation, before attenuation, and are representative of the mixing photon-number statistical properties. The internal quantum conversion efficiencies are computed in terms of the mixing depletion as $\eta = \left(1 - \langle n_{\mathrm{M}_0} \rangle / \langle n_{\mathrm{M}_0,\mathrm{off}} \rangle\right) \times 100$ to make the measurement agnostic to the linear propagation losses of the fiber, where $\langle n_{\mathrm{M}_0} \rangle$ and $\langle n_{\mathrm{M}_0,\mathrm{off}} \rangle$ are the expected photon numbers of the mixing frequency measured at the fiber output when the conversion is respectively on and off. We obtain up-conversion efficiencies reaching 70% and down-conversion efficiencies of almost 90%, in good agreement with the simulations displayed in Fig. 4a. We attribute the difference in the attainable up- and down-conversion efficiencies to the pressure-dependent spatio-temporal dynamics of the Raman coherence along the fiber (see Supplementary Information). The results of spectral measurements are shown for both transduction processes in Fig. 4c, where the counts are normalized relative to the mixing counts when the transduction is turned off. We observe that no parasitic sidebands related to the mixing frequency appear, which indicates the selectivity of the phase-matching conditions. The narrowband character of the spectral lines facilitates the filtering and selection of the frequencies of interest.

**Preservation of statistical properties**

After each transduction efficiency measurement, which are displayed in Fig. 4b, we carried out an analysis of the transfer of photon number statistics from the original mixing frequency to the converted counterpart. The output frequencies of interest are spectrally filtered for photodetection. In every measurement, we collect 20000 samples in order to have statistical significance, with each one of these samples corresponding to an individual pulse of the selected frequency. From these samples, we obtain histograms corresponding to the mixing photon statistics when the conversion is turned off and the resulting converted statistics when the conversion is on, with each case grouped in independent subplots of Fig. 5. Figs. 5b and 5c magnify two cases as example, while overlaying both the original and converted statistics.

The original mixing statistics are depicted in orange, while the up- and down-converted statistics are depicted in blue and red color respectively. Each dot represents the top of a histogram bar and the solid curves correspond to fits using Eq. (1). In addition to the histograms, we have also presented box plots to facilitate a meaningful visual comparison of the results. To this end, the photon count axis has been appropriately normalized relative to the corresponding mean value of each histogram. The qualitative similarity of all pairs of "on-off" statistical distributions displayed in Fig. 5 is evident. However, in order to provide quantitative estimates for the resemblance of statistics, we consider two metrics, which are detailed in Materials and Methods: the ratio of zero-delay second-order correlation functions $g^{(2)}_{\mathrm{M}_l}(0)$, with $l = -1, 0, 1$ labeling the down-converted, mixing and up-converted frequencies, and the fidelity $\mathcal{F}$ between two states. Table 1 presents the retrieved values of the two metrics considered, both for frequency up- and down-conversion. The close-to-unity values unambiguously corroborate the ability of molecular modulation-based frequency transduction to successfully preserve the photon-number statistics.

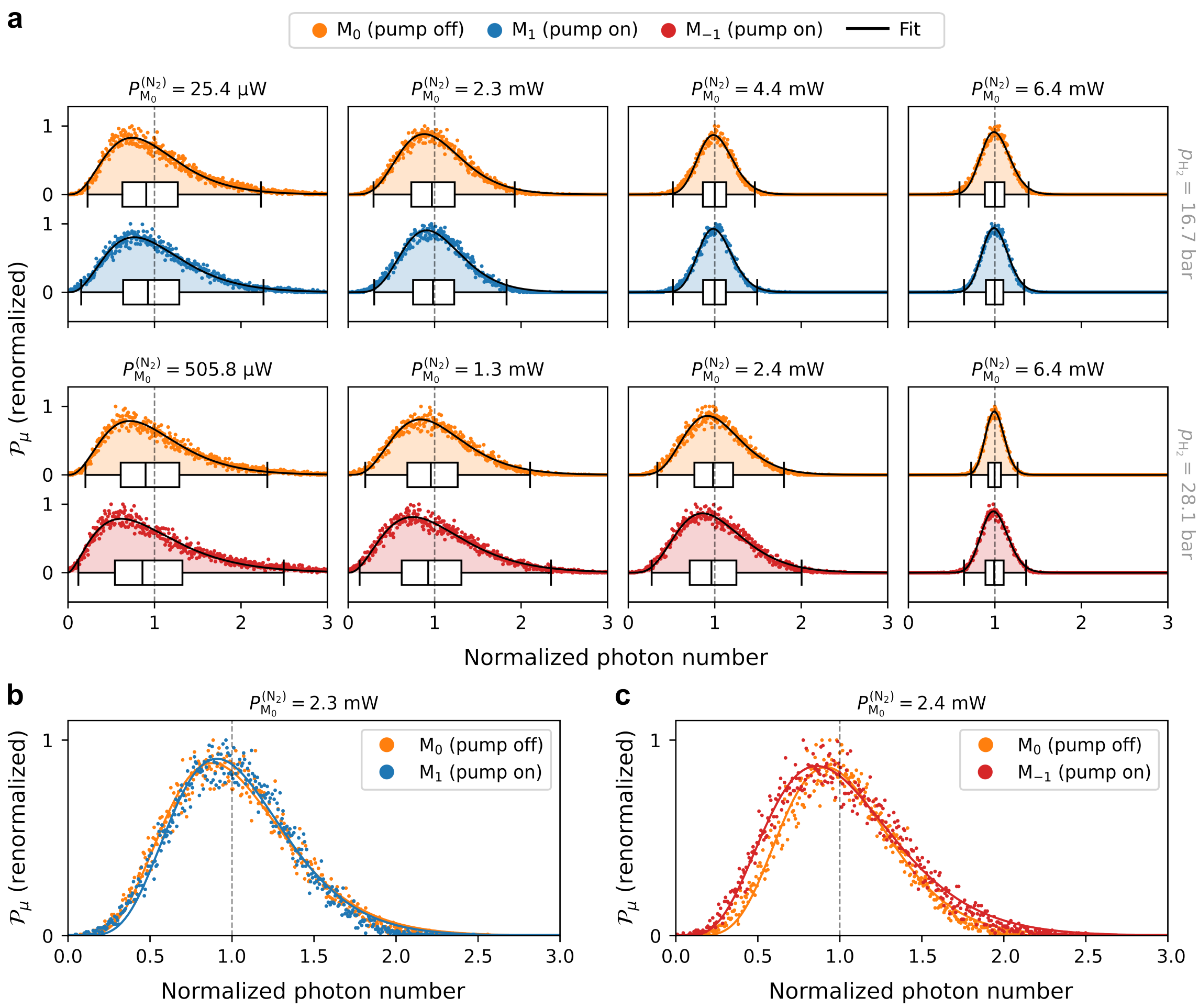

**Fig. 5** Photon statistics for the conversion processes. **a** The top row of subplots displays the up-conversion results, while the bottom row corresponds to down-conversion measurements. Each subplot shows the measurement of the mixing statistics when the transduction is turned off and the converted frequency statistics. The histograms are fitted to gamma distributions, the photon number axis is normalized with respect to the mean of each histogram and the probabilities are renormalized to their maximum value. Box plots are shown below each histogram, indicating the quartiles and median of the distributions. **b**-**c** Zoom into two of the cases shown at **a** to facilitate the visualization and comparison of the resulting histograms.

**Table 1** Transduction quantum statistical comparison metrics

| Process | $P_{\mathrm{M_0}}^{(\mathrm{N_2})}$ | $g_{\mathrm{M_{\pm1}}}^{(2)}(0)/g_{\mathrm{M_0}}^{(2)}(0)$ | $\mathcal{F}$ |
|---|---|---|---|
| Up-conversion | 25.4 μW | 1.009 ± 0.007 | 0.9962 ± 0.0007 |
| | 2.3 mW | 0.983 ± 0.004 | 0.9972 ± 0.0006 |
| | 4.4 mW | 0.998 ± 0.002 | 0.9997 ± 0.0002 |
| | 6.4 mW | 0.9946 ± 0.0013 | 0.9934 ± 0.0009 |
| Down-conversion | 505.8 μW | 1.081 ± 0.009 | 0.9863 ± 0.0015 |
| | 1.3 mW | 1.073 ± 0.006 | 0.9835 ± 0.0016 |
| | 2.4 mW | 1.048 ± 0.004 | 0.9814 ± 0.0018 |
| | 6.4 mW | 1.124 ± 0.0009 | 0.938 ± 0.002 |

## 3 DISCUSSION

We studied the capacity of molecular modulation-based frequency transduction to keep the statistical fluctuations of the incoming radiation unaltered upon efficient conversion. For this purpose, we developed a transducer consisting of a hydrogen-filled HCF with carefully engineered dispersion. This property determines the spatial beating of the molecular coherence wave driven by the pump pulse and its down-shifted Stokes component and, therefore, establishes the phase-matching conditions for the transduction of weak mixing pulses with intensities below the Raman threshold of the system. Simulations based on the Maxwell-Bloch formalism reveal that the transducer can be set for either frequency up- or down-shifting by adequately tuning the gas pressurization to ~16.7 or ~28.1 bar without any crosstalk between the two processes. This is verified by experimental observations, confirming the high internal quantum efficiencies predicted with measured up-conversion values reaching 70% and approaching 90% for down-conversion. It is important to remark that the internal efficiency is the most adequate metric to assess the performance of the transductor solely on the basis of the underlying physics, since it is largely independent of the losses of a specific setup. However, for completeness we have also measured the external quantum efficiency as the ratio of the outcoupled photon numbers for the mixing sidebands to those of the mixing signal at the fiber input. In the current system we have obtained external efficiencies of ~30% for down-conversion and ~47% for up-conversion, that could be improved in the future using new designs of ultrathin-walled HCFs featuring lower attenuation[28].

The photon statistics of both the mixing and converted frequencies were measured. To characterize these results, we computed the ratio of zero-delay second-order correlation functions and the fidelity between the photon-number distributions of both photonic states. Despite the slight increase in the normalized variance of the $\mathrm{M_{-1}}$ band (see Fig. 5c), that we attribute to higher loss and reduced signal-to-noise ratio, the near-unity values obtained indicate a good degree of preservation of the normalized statistical properties during molecular modulation. This aligns with the theoretical expectations for this phenomenology, as the molecular modulation-based transducer should behave as a frequency beam splitter in which the efficiency of the conversion does not depend on the intensity of the mixing light if operated below the Raman threshold[9, 37]. In this regard, this beam-splitter behavior of frequency transducers strictly requires near-unit transduction efficiencies to faithfully transfer the photon-number properties of the incoming fields, which molecular modulation is able to achieve[37].

For this proof-of-concept experiment, the fiber length was of $3.7 \pm 0.1$ m, which allows to reduce the pump energies required for optimal molecular modulation by two orders of magnitude compared to previous experiments[32], opening the door to the future use of modest pump lasers with a very small footprint. Furthermore, by driving the molecular coherence using visible light, this phenomenology is observed at much lower pressures than for the case of near-infrared pumping (e.g. optimal up-conversion occurs at ~16.7 bar against the 70 bar reported in ref. 32). Interestingly, operation at lower pressures increases the dephasing time of the hydrogen molecules, which relaxes the synchronization requisites between the pump and the mixing pulses. The size of the transduction stage can be further optimized in future works by combining the hydrogen gas with noble gases to scale-down the dynamics[46].

With these results, we demonstrate the potential of molecular modulation as an effective method for bridging disparate spectral regions by using existing fiber-compatible technology. The ability to preserve both quantum correlations and photon-number statistics is of special interest for quantum information processing, including quantum key distribution using decoy states[47]. While the selective shift of $\pm 125$ THz in $H_2$ might enable the efficient interconnection of devices that operate in the visible with those operating in the near-infrared, this phenomenology is not limited to these wavelengths or frequency shift. The tunable phase-matching properties of HCF-based molecular modulation and its ability to operate broadband[34], even with other gases providing different frequency shifts, open the door to novel fiber-based transducers optimized for operation in otherwise hardly accessible spectral regions such as the ultraviolet or the mid-infrared.

# 4 MATERIALS AND METHODS

**Fiber-based light source with tunable photon-number statistics**

To gain access to a source of yellow light capable of providing different photon-number statistics, we built a custom Raman converter consisting of a tubular HCF pressurized with $N_2$ gas at $25.26 \pm 0.01$ bar pumped at a repetition rate of 1 kHz by green 515 nm pulses having a full width at half maximum duration (FWHM) of $1.42 \pm 0.03$ ns. The fiber, with core diameter of ~21 μm and average capillary-wall thickness of ~470 nm, was cut to a length of $0.71 \pm 0.01$ m. It can guide the wavelengths of interest with losses of $\alpha_{515} = 0.057$ dB/m and $\alpha_{585} = 0.041$ dB/m, where the subscripts denote the corresponding wavelength. At 515 nm, we routinely achieve coupling efficiencies exceeding 94% into the $LP_{01}$-like fundamental guided mode of the fiber. By careful adjustment of the average power of the pump light, we retrieve 585 nm pulses in the form of $LP_{01}$-like Stokes radiation. The SRS process involved here is given by the fundamental Q(0) vibrational transition of molecular nitrogen, which provides a Raman shift of approximately 70 THz. The FWHM time duration of the resulting Stokes pulses is of $0.3 \pm 0.1$ ns according to simulations shown at Fig. 2b with a 1.42 ns-long pump and varies slightly within the given uncertainty depending on the pump power.

## Maxwell-Bloch description of molecular modulation

SRS in gas-filled HCFs can be described by the following set of Maxwell-Bloch equations[24, 35, 48]:

$$\frac{\partial E_l}{\partial z} = -\frac{1}{2}\alpha_l E_l - i\kappa_{2,l}\frac{\omega_l}{\omega_{l-1}} Q E_{l-1} e^{i(\beta_l-\beta_{l-1})z} - i\kappa_{2,l+1} Q^* E_{l+1} e^{-i(\beta_{l+1}-\beta_l)z}, \quad (2)$$

$$\frac{\partial Q}{\partial t} = -\frac{Q}{T_2} + \frac{i}{4} w \sum_l \kappa_{1,l} E_l E_{l-1}^* e^{-i(\beta_l-\beta_{l-1})z}. \quad (3)$$

These equations are written for single-mode linearly co-polarized fields propagating along the $z$ direction with envelopes $E_l(z,t)$, where $l$ denotes a given Raman sideband with propagation constant $\beta_l$ and angular frequency $\omega_l = \omega_\mathrm{P} + l\Omega_R$ with $\Omega_R$ being the Raman shift, $\omega_\mathrm{P}$ being the pump frequency and $l \in \mathbb{Z}$. $Q(z,t)$ denotes the amplitude of the coherence wave induced by the contributions of all the consecutive sideband pairs. Here, $\kappa_{1,l}$ and $\kappa_{2,l}$ are coupling coefficients, $\alpha_l$ is the linear propagation loss of the corresponding sideband and $T_2$ is the molecular dephasing time. Finally, $w$ denotes the molecular population inversion, usually set to $w \approx -1$ assuming that most of the molecules are not excited during the scattering.

To describe molecular modulation, we need to consider introducing another field of frequency $\omega_{\mathrm{M}_0}$ that will play the role of mixing signal. This field remains below the Raman threshold under all conditions, and is therefore unable to contribute to the coherence $Q$. However, the mixing will be coupled to the pre-existing coherence and will therefore be able to generate its own sidebands, $E_m(z,t)$, with frequencies $\omega_m = \omega_{\mathrm{M}_0} + m\Omega_R$, where $m \in \mathbb{Z}$. Therefore, to describe this effectively-linear interaction, the following set of coupled equations is also considered in addition to the previous one[48]:

$$\frac{\partial E_m}{\partial z} = -\frac{1}{2}\alpha_m E_m - i\kappa_{2,m}\frac{\omega_m}{\omega_{m-1}} Q E_{m-1} e^{i(\beta_m-\beta_{m-1})z} - i\kappa_{2,m+1} Q^* E_{m+1} e^{-i(\beta_{m+1}-\beta_m)z}. \quad (4)$$

If we consider that the pump field $E_\mathrm{P}$ and its Stokes sideband $E_{\mathrm{S}_1}$ are the main contributors to the coherence $Q$, the mixing sideband pair with a propagation constant difference $\beta_m - \beta_{m-1}$ that matches the beating $\beta_\mathrm{P} - \beta_{\mathrm{S}_1}$ of the coherence wave will exchange photon population more efficiently than the rest of the sidebands. The propagation constants are governed by the dispersion of the fiber-gas system. Therefore, for a given pump wavelength, the mixing frequency to be shifted and whether this shift is positive or negative can be selected by both an adequate design of the HCF cross-section and judicious adjustment of the filling gas pressure.

To simulate the nonlinear dynamics of the fields propagating along the gas-filled HCF, Eqs. (2)-(3) and (4) are numerically solved using a fourth-order Runge-Kutta scheme including an adaptive propagation step.

## Experimental setup

In the setup, depicted at Fig. 2a, both fibers are pumped by linearly-polarized 515 nm pulses of $1.42 \pm 0.03$ ns FWHM at a repetition rate of 1 kHz. These pulses are obtained by frequency doubling a 1030 nm Q-switched diode-pumped solid-state laser in a Lithium Triborate (LBO) nonlinear crystal. To feed the two HCFs, the main stream of pulses is split by a balanced beam splitter into two paths, each one having an independent power control consisting of a half-wave plate and a polarizing beam splitter. A free-space delay line is introduced to compensate the path difference and synchronize the pulses before reaching the hydrogen-filled HCF. Each HCF used is placed inside an arrangement of two tube-connected gas cells sealed by magnesium fluoride windows and pressurized according to the needs of the experiment.

The HCF chosen as transduction platform for the experiment features a kagome cross-sectional geometry, with a flat-to-flat diameter $d_{\text{ff}} \sim 20.7$ µm and a core-wall thickness of ~120 nm. The main guidance window in this fiber is broadband enough to avoid loss-inducing resonances close to the wavelengths of interest, exhibiting attenuation of $\alpha_{515} = 0.53$ dB/m, $\alpha_{585} = 0.61$ dB/m, $\alpha_{655} = 0.67$ dB/m and $\alpha_{773} = 1.78$ dB/m with subscripts indicating the wavelengths. The attenuation of the $M_1$ sideband could not be experimentally measured since this wavelength lies outside the spectral coverage of the supercontinuum source used for characterization. However, based on the agreement with numerical modelling, we estimated the loss coefficient of this band to be $\alpha_{471} \simeq 0.65$ dB/m, a reasonable value taking into account the tendency of the experimental loss curve (see Supplementary Information). The kagome HCF, with a length of $3.7 \pm 0.1$ m, transmits approximately 70% of the 515 nm and 585 nm power, propagating the light in the $LP_{01}$-like mode. For HCFs with this type of geometry, the dispersion can be well approximated within guidance windows by the capillary dispersion model[49]. In this approximation, the effective capillary diameter that preserves the original kagome core area is $d_{\text{ap}} \approx 1.05\, d_{\text{ff}}$. Considering this modeling of the dispersion, and confirmed by experimental measurements with the hydrogen-filled kagome HCF, we determine up- and down-conversion phase-matching pressures of approximately 16.7 bar and 28.1 bar for a 515 nm pump and a 585 nm mixing.

Regarding pump energies, we have found that the up-conversion (down-conversion) dynamics is optimized for ~0.5 µJ (~1.5 µJ), in good agreement with the simulated values (see Supplementary Information for details). To quantify the internal quantum efficiency $\eta$, the detection is carried out by a silicon photodiode placed inside an optical enclosure at the fiber output to reduce the influence of ambient light and the mixing frequency is selected using a bandpass filter. For each case, the detector is zeroed before each measurement to reduce inaccuracies.

## Spectral characterization

The analysis of the spectral content of the signals involved in the transduction was carried out by diverting the radiation exiting the $H_2$-filled HCF to an integrating sphere connected to a CCD spectrometer. The device is not intensity-calibrated, and therefore the different spectral intensities cannot be quantitatively compared to each other. The complete spectrum is obtained by optimizing the conversion process through adjustment of the pump power and settings in the spectrometer, so that the pump wavelength peak is close to saturation. Before taking the spectral measurement, we perform background subtraction in order to clean ambient noise. After a complete measurement round, the pump is blocked to permit the characterization of the mixing peak in absence of other Raman sidebands. Lastly, we use the measured standalone-mixing peak counts from the spectrometer to normalize the full spectrum of the conversion.

**Photodetection statistical measurements**

The photon-number distributions are measured using two silicon-based photodiode detectors, placed equidistantly as depicted in Fig. 2a. One of the detectors collects the residual pump light and is used to trigger the detection at the second photodiode, which is used to collect the mixing or converted light. Both photodetectors are connected to an oscilloscope, with which we are able to send the time-dependent voltage response signal to a computer for storage and processing. Then, we integrate this signal with respect to time and convert it to photon number values through a calibration factor. The photodetector collecting the mixing or converted pulses is placed inside an optical enclosure to reduce the effects of residual ambient light. With this, the resulting photon-number histogram of dark counts is symmetrical and centered at zero with a standard deviation two orders of magnitude lower than the signals detected during transduction measurements.

**Metrics for the quantification of statistical similarity**

The values of the correlation functions $g^{(2)}(0)$ are retrieved from the fitted mean and variance of the photon-number probability distributions. The general formula used is

$$g^{(2)}(0) = 1 + \frac{\langle(\Delta n)^2\rangle - \langle n\rangle}{\langle n\rangle^2}, \tag{5}$$

where $\langle(\Delta n)^2\rangle$ is the expected variance and $\langle n\rangle$ the mean photon number. The gamma distribution Eq. (2) presents the mean $\langle n\rangle$, while the variance reads as $\langle(\Delta n)^2\rangle = \langle n\rangle^2/\mu$. Note that in the limit of large average photon number, the resulting $g^{(2)}(0)$ value becomes heavily dependent on $\mu$. The fidelity between two quantum states is a useful metric to compare two states characterized by density matrices. Specifically, if the two density matrices involved are diagonal such as in the case of two thermal states, the fidelity can be written as[50]

$$\mathcal{F} = \left(\mathrm{tr}\sqrt{\sqrt{\rho}\sigma\sqrt{\rho}}\right)^2 = \left(\sum_i \sqrt{r_i s_i}\right)^2. \tag{6}$$

Here $\rho$ and $\sigma$ would be the diagonal density matrices with elements $r_i$ and $s_i$ respectively, where tr indicates the trace operation. The values of this metric range from 0 for completely different states to 1 for states with the same density matrix. To use this metric, the photon statistics in Fig. 5 are considered maintaining the normalized photon number axis. This is motivated by the gamma distribution formula Eq. (2) presenting the photon number variable always normalized as $n/\langle n\rangle$. Under these conditions, we take the histogram values as the diagonal elements of the density matrices with normalized photon numbers.

## DECLARATIONS

### Acknowledgements

This work was supported by the grants PID2024-158310NB-I00, PID2024-155582OB-C31, CNS2025-166590 and PRE2022-102843, funded by MICIU/AEI/10.13039/501100011033, by ERDF, EU, and "ESF+", by Gobierno Vasco/Eusko Jaurlaritza under Grant IT1829-26, ELKARTEK SmartµS 2025/00058 and Newhegaz KK 2025/00074, by “Translight” Initiative of EHU, and by the IKUR Strategy of the Department of Education of the Basque Government under Grant IKUR-IKA-23/03. MS-H acknowledges support from the predoctoral grant “Formación de Profesorado Universitario” FPU22/01451 from the Spanish Ministry of Science, Innovation and Universities (MICIU).

### Author contributions

D.N. conceived and supervised the project. A.M., P.A. and D.N. designed the experiment. A.M. implemented the two-stage setup and performed the measurements and simulations, with assistance from P.A. and M.S.-H. The specialty HCFs were provided by F.B., F.G., B.D. and N.J. J.Z. and D.N. provided funding support. A.M. and D.N. curated the data and wrote the initial draft. All authors contributed to the discussion and revision of the manuscript.

### Data availability

All data supporting the main results of this work are available from the corresponding authors upon reasonable request.

### Conflict of interest

The authors declare no competing interests.

## REFERENCES

1. Awschalom, D. *et al*. Development of Quantum Interconnects (QuICs) for Next-Generation Information Technologies. *PRX Quantum* **2**, 017002 (2021).

2. Knörzer, J., Liu, X., Schiffer, B. & Tura Brugués, J. Distributed quantum information processing: a review of recent progress. *Reports on Progress in Physics* (2026).

3. Somaweera, D. *et al*. Rydberg Atom-Based Sensors: Principles, Recent Advances, and Applications. *Photonics* **12** (2025).

4. Kumar, V., Cicconetti, C., Conti, M. & Passarella, A. Quantum internet: Technologies, protocols, and research challenges. *International Journal of Networked and Distributed Computing* **13**, 22 (2025).

5. Tanzilli, S. *et al*. A photonic quantum information interface. *Nature* **437**, 116–1120 (2005).

6. Murphy, L. R. *et al*. Tunable frequency conversion in doped photonic crystal fiber pumped near degeneracy. *Optica* **11**, 1490–1496 (2024).

7. Fan. L. *et al*. Integrated optomechanical single-photon frequency shifter. *Nature Photonics* **10**, 766–770 (2016).

8. Lauk, N. *et al*. Perspectives on quantum transduction. *Quantum Science and Technology* **5**, 020501 (2020).

9. Wang, C., Li, F. & Jiang, L. Quantum capacities of transducers. *Nature Communications* **13**, 6698 (2022).

10. Epple, G. *et al*. Rydberg atoms in hollow-core photonic crystal fibres. *Nature communications* **5**, 4132 (2014).

11. Langbecker, M., Noaman, M., Kj aergaard, N., Benabid, F. & Windpassinger, P. Rydberg excitation of cold atoms inside a hollow-core fiber. *Phys. Rev. , A* **96**, 041402(R) (2017).

12. Xin, M., Leong, W. S., Chen, Z. & Lan, S. Transporting Long-Lived Quantum Spin Coherence in a Photonic Crystal Fiber. *Phys.Rev.Lett.* **122**, 163901 (2019).

13. Leong, W. S. *et al*. Large array of Schrödinger cat states facilitated by an optical waveguide. *Nature communications* **11**, 5295 (2020).

14. Wang, R. *et al*. Optical trapping of mesoscale particles and atoms in hollow-core optical fibers: principle and applications. *Light Sci Appl* **14**, 146 (2025).

15. Russell, P. S. J., H olzer, P., Chang, W., Abdolvand, A. & Travers, J. C. Hollow-core photonic crystal fibres for gas-based nonlinear optics. *Nature Photonics* **8**, 278–286 (2014).

16. Petrovich, M. *et al*. Broadband optical fibre with an attenuation lower than 0.1 decibel per kilometre. *Nat. Photon* **19**, 1203–1208 (2025).

17. Gao, S. *et al*. Fourfold truncated double-nested antiresonant hollow-core fiber with ultralow loss and ultrahigh mode purity. *Optica* **12**, 56–61 (2025).

18. Ding, Y. *et al*. Support Tube DNANF Achieving 0.051 dB/km and Scaled to > 10000 km of Production. *J. Lightwave Technol.* **44**, 7085–7089 (2026).

19. Sprague, M. R. *et al*. Broadband single-photon-level memory in a hollow-core photonic crystal fibre. *Nature Photonics* **8**, 287–291 (2014).

20. McGarry, C. *et al*. Low-loss, compact, fibre-integrated cell for quantum memories. *Opt.Express* **32**, 21925–21935 (2024).

21. Clark, M. J. *et al*. Coexistence of entanglement-based quantum channels with DWDM classical channels over hollow core fibre in a four node quantum communication network. *npj Quantum Inf* **11**, 181 (2025).

22. Carosini, L. *et al*. Quantum communication with quantum dots beyond telecom wavelengths via hollow-core fibers. *Optica Quantum* **4**, 82–88 (2026).

23. Benabid, F., Knight, J. C., Antonopoulos, G. & Russell, P. S. J. Stimulated Raman scattering in hydrogen-filled hollow-core photonic crystal fiber. *Science* **298**, 399–402 (2002).

24. Arcos, P. *et al*. Narrowband stimulated Raman scattering and molecular modulation in anti-resonant hollow-core fibres. *Europhys. Lett.* **147**, 45001 (2024).

25. McGarry, C., Harrington, K., Davis, A. O. C., Mosley, P. J. & Rusimova, K. R. Microstructured optical fibers for quantum applications: Perspective. *APL Quantum* **1**, 030901 (2024).

26. Wang, Z. *et al*. Hollow-core fiber gas lasers. *Light Sci Appl* **15**, 208 (2026).

27. Elu, U. *et al*. Seven-octave high-brightness and carrier-envelope-phase-stable light source. *Nature Photonics* **15**, 277–280 (2021).

28. Sabbah, M. *et al*. Resonance-free deep ultraviolet to near infrared supercontinuum generation in a hollow-core antiresonant fibre. *Journal of Physics: Photonics* **7**, 025019 (2025).

29. Finger, M. A., Joly, N. Y., Russell, P. S. J. & Chekhova, M. V. Characterization and shaping of the time-frequency Schmidt mode spectrum of bright twin beams generated in gas-filled hollow-core photonic crystal fibers. *Physical Review A* **95**, 053814 (2017).

30. Cordier, M., Delaye, P., Gérôme, F., Benabid, F. & Zaquine, I. Raman-free fibered photon-pair source. *Scientific reports* **10**, 1650 (2020).

31. Lopez-Huidobro, S., Noureddin, M., Chekhova, M. V. & Joly, N. Y. Tunable fiber source of entangled UV-C and infrared photons. *Opt. Lett.* **48**, 3423–3426 (2023).

32. Tyumenev, R., Hammer, J., Joly, N. Y., Russell, P. S. J. & Novoa, D. Tunable and state-preserving frequency conversion of single photons in hydrogen. *Science* **376**, 621–624 (2022).

33. Hamer, A., Vewinger, F., Peters, T., Frosz, M. H. & Stellmer, S. Frequency conversion in a hydrogen-filled hollow-core fiber using continuous-wave fields. *Opt.Lett.* **49**, 6952–6955 (2024).

34. Bauerschmidt, S. T., Novoa, D., Abdolvand, A. & Russell, P. S. J. Broadband-tunable LP01 mode frequency shifting by Raman coherence waves in a H2-filled hollow-core photonic crystal fiber. *Optica* **2**, 536–539 (2015).

35. Raymer, M. G. & Walmsley, I. A. III The quantum coherence properties of stimulated Raman scattering. **28**, 181–270 (1990).

36. Chen, Y. & Wise, F. Unified and vector theory of Raman scattering in gas-filled hollow-core fiber across temporal regimes. *APL photonics* **9** (2024).

37. Gonzalez-Raya, T. *et al*. Entanglement transfer during quantum frequency conversion in gas-filled hollow-core fibers. *APL Photonics* **10** (2025).

38. Wang, J., Sokolov, A. V. & Agarwal, G. S. Quantum entanglement between signal and frequency-up-converted idler photons. *Physical Review A* **108**, 063706 (2023).

39. Dynes, J. F. *et al*. Testing the photon-number statistics of a quantum key distribution light source. *Optics express* **26**, 22733–22749 (2018).

40. Foletto, G., Picciariello, F., Agnesi, C., Villoresi, P. & Vallone, G. Security bounds for decoy-state quantum key distribution with arbitrary photon-number statistics. *Phys. Rev. , A* **105**, 012603 (2022).

41. Thalassinos, G. *et al*. Robust quantification of the diamond nitrogen-vacancy center charge state via photoluminescence spectroscopy. *APL Photonics* **10** (2025).

42. Heller, L., Lowinski, J., Theophilo, K., Padrón-Brito, A. & de Riedmatten, H. Raman storage of quasideterministic single photons generated by Rydberg collective excitations in a low-noise quantum memory. *Physical Review Applied* **18**, 024036 (2022).

43. Kolyadin, A. N. *et al*. Light transmission in negative curvature hollow core fiber in extremely high material loss region. *Optics express* **21**, 9514–9519 (2013).

44. Mehta, C. L. VIII Theory of Photoelectron Counting. *Progress in Optics* **8**, 373–440 (1970).

45. Nattermann, K., Fabricius, N. & Von Der Linde, D. Observation of transverse effects on quantum fluctuations in stimulated Raman scattering. *Opt. Commun.* **57**, 212–216 (1986).

46. Arcos, P. *et al*. Scaling of nonlinear dynamics driven by stimulated Raman scattering in gas-filled hollow-core fibers. *J. Lightwave Technol.* **43**, 8406–8414 (2025).

47. Lo, H., Ma, X. & Chen, K. Decoy State Quantum Key Distribution. *Phys.Rev.Lett.* **94**, 230504 (2005).

48. Bauerschmidt, S. in *Parametric control of raman scattering in hollow-core photonic crystal fiber* (Erlangen Scientific Press, Friedrich-Alexander-Universität Erlangen-Nürnberg, 2016).

49. Finger, M. A., Joly, N. Y., Weiss, T. & Russell, P. S. J. Accuracy of the capillary approximation for gas-filled kagomé-style photonic crystal fibers. *Opt. Lett.* **39**, 821–824 (2014).

50. Nielsen, M. A. & Chuang, I. L. in *Quantum Computation and Quantum Information: 10th Anniversary Edition* (Cambridge University Press, Cambridge, 2010).

Supplementary Information

# Preservation of photon statistics upon selective molecular modulation in hollow-core fibers

Arturo Mena[1*], Pau Arcos[1], María Sánchez-Hernández[1], Benoit Debord[2], Frédéric Gérôme[2], Fetah Benabid[2], Nicolas Y. Joly[3,4], Oskar Arrizabalaga[5], Joseba Zubia[1,6], David Novoa[1,6,7*]

[1]Department of Communications Engineering, University of the Basque Country (EHU), Bilbao, 48013, Spain.
[2]GPPMM Group, XLIM Institute, CNRS UMR 7252, University of Limoges, Limoges, 87100, France.
[3]Department of Physics, University of Erlangen-Nürnberg, Erlangen, 91058, Germany.
[4]Max Planck Institute for the Science of Light, Erlangen, 91058, Germany.
[5]Department of Applied Mathematics, University of the Basque Country (EHU), Bilbao, 48013, Spain.
[6]EHU Quantum Center, University of the Basque Country (EHU), Bilbao, Spain.
[7]IKERBASQUE, Basque Foundation for Science, Bilbao, 48009, Spain.

*Corresponding author(s). E-mail(s): arturo.mena@ehu.eus; david.novoa@ehu.eus;

## Contents

# 1 Details of the experimental setup

Fig. 1 displays an expanded depiction of the actual experimental setup as implemented in the laboratory. Fig. 1a shows the main linearly-polarized 515 nm beam being split by a balanced beam splitter into the two main paths of the setup to pump the two pressurized hollow-core fibers (HCF). Each path is equipped with a power control consisting of a half-wave plate and a polarizing beam splitter in order to adjust the pulse energies according to the needs of each fiber stage. To compensate the optical path difference, we introduce a free-space delay line built with a dovetail rail and a roof mirror. The pump of the $H_2$-filled HCF is combined at a dichroic mirror with the 585 nm signal coming from the $N_2$-filled HCF. In the experiment, the $N_2$-filled HCF is pumped at 1 kHz with average 515 nm powers in the range of 26 to 31 mW as seen in Fig. 2b of the main text, while in the $H_2$-filled HCF the pump power lays within 0.5 to 1.5 mW to optimize the transduction efficiencies depending on whether up- or down-conversion is selected.

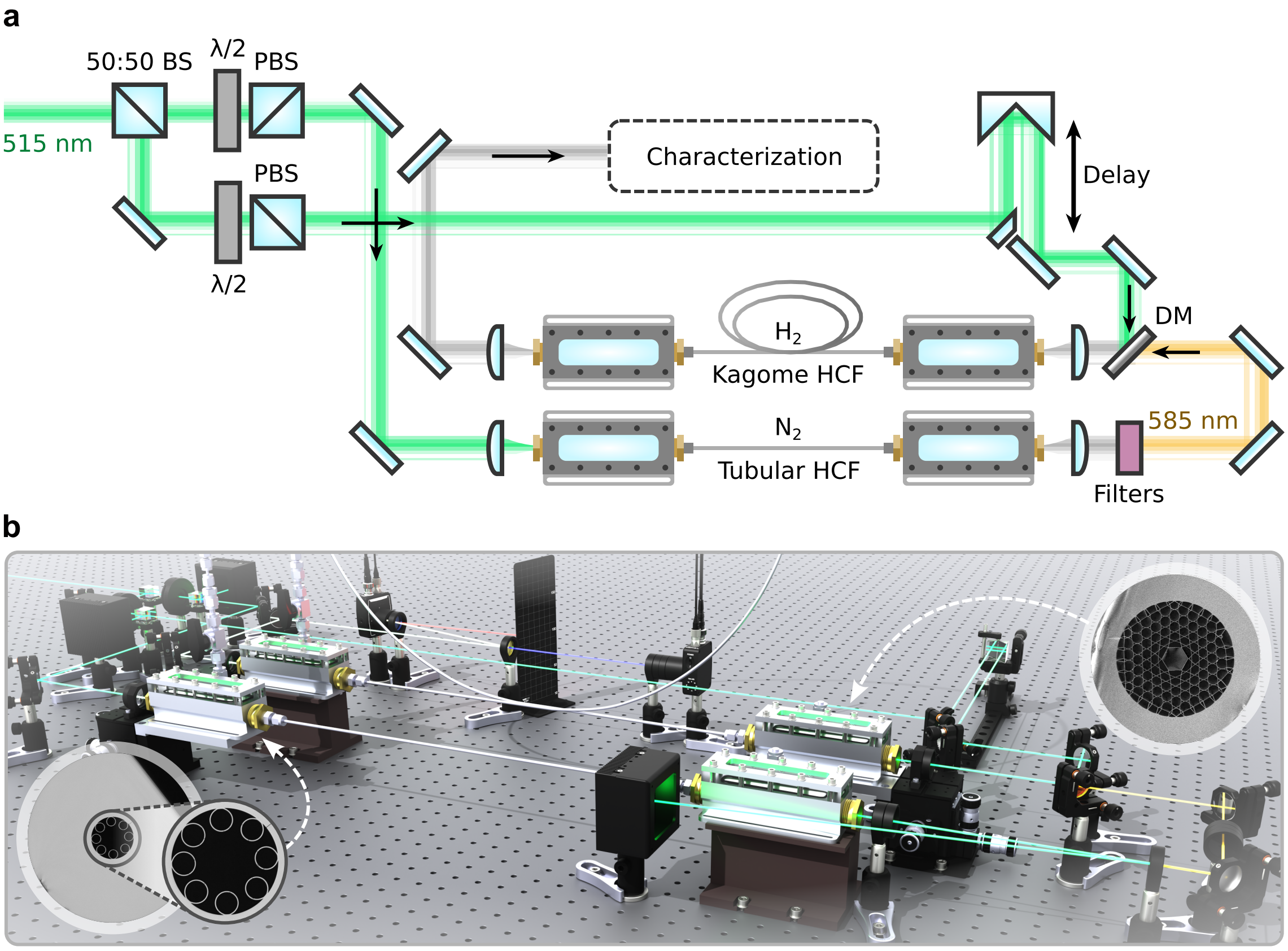

**Fig. 1** Expanded depiction of the experimental setup. **a** Sketch of the setup as implemented in the laboratory. Here, BS denotes a balanced beam splitter, $\lambda/2$ half-wave plates, PBS polarizing beam splitters and DM a dichroic mirror. **b** Illustrative 1:1 scale depiction of the setup with no optical enclosures placed to show all the elements.

Fig. 1b represents the setup in 1:1 scale as implemented in the lab. The cross-sections of the HCF are also shown, indicating the corresponding pair of tube-connected gas cells in which they are placed. The guidance losses of these HCFs measured using the standard cut-back technique are shown in Fig. 2. The up-converted wavelength of ~471 nm is not included in Fig. 2b since it is out of the range allowed by the supercontinuum light source used for characterization.

Regarding the quantification of the photon statistics, the wavelength-specific calibration factor used to convert the voltage signals from photodetection to photon number values is computed beforehand by using the same frequencies to be detected. To obtain this factor, we turn off the transduction pump and increase the power of the mixing pulses to be well above the Raman threshold of the $H_2$-filled kagome HCF so that all the required wavelengths can be obtained as SRS sidebands of the mixing signal in a regime where the energy fluctuation statistics are mean-symmetrical and with low variance. With this, the average power can be measured for each wavelength with reduced uncertainty using spectral bandpass filters and a silicon-based photodiode power sensor. Therefore, considering the laser repetition rate, the average energy per pulse can be obtained and converted to number of photons to retrieve the calibration factor that converts the integrated voltage signal to photon number.

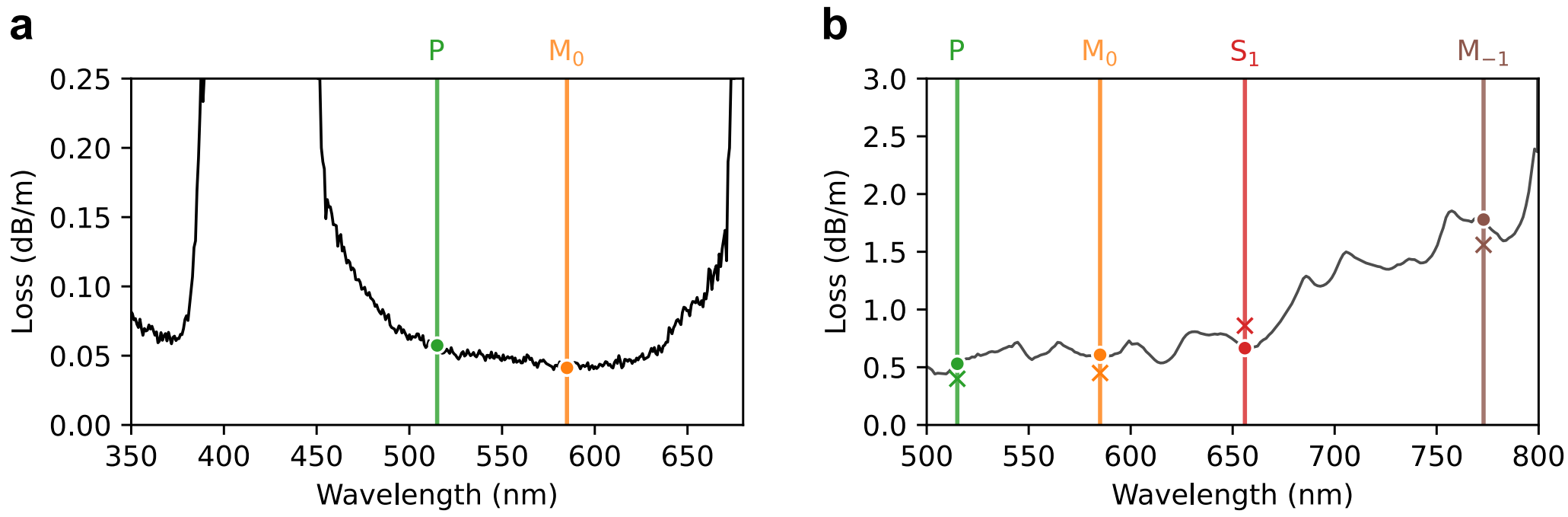

**Fig. 2** Guidance characterization of the HCFs used in the experiment. **a** Losses of the tubular HCF used for mixing state generation. **b** Losses of the kagome HCF used as frequency transducer. Both in **a** and **b**, the colored vertical lines indicate the wavelengths of interest in each fiber. The loss values corresponding to characterization data are marked with circles, while the slightly different values used in the simulations to best fit the experimental phenomenology are marked with crosses.

## 2 Spatial selection of wavelengths

The frequencies involved in the experiment are sufficiently narrowband and spectrally distant as to allow easy spectral selection. This can be done by, e.g., using bandpass filters or dispersive elements like a prism. For the latter example, Fig. 3 shows photographies of the collimated profiles exiting the transduction system. To obtain these images, the signals are first collimated, then separated spatially by a prism and, finally, cast onto a flat white screen.

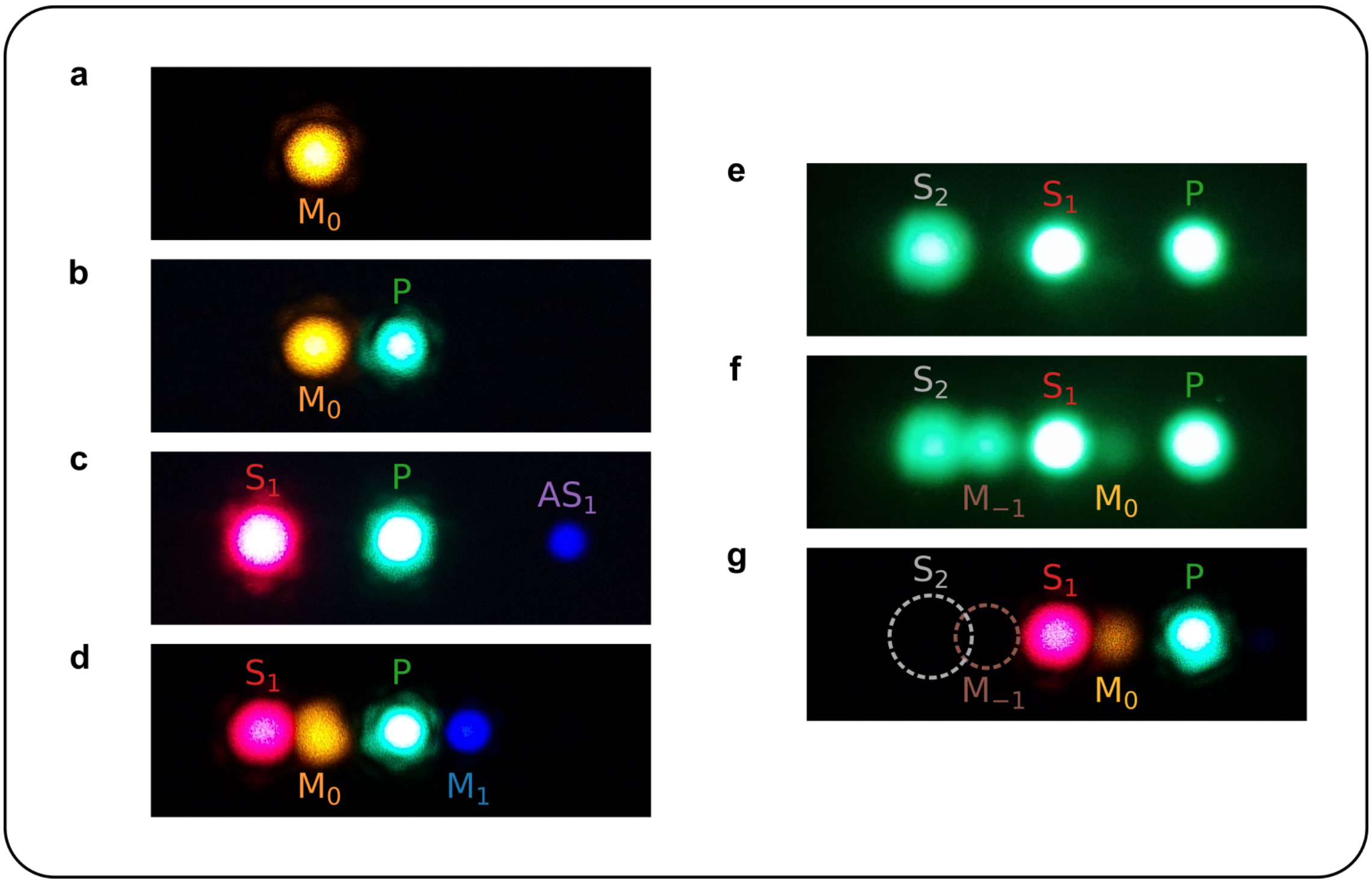

**Fig. 3** Photographies of the collimated mode profiles of the different wavelengths exiting the transduction system. **a** Collimated mixing ($M_0$) beam in the case when this wavelength is the only one injected into the kagome HCF. **b** Mixing and pump (P) profiles when both beams are below the Raman threshold and the transduction is consequently inactive. **c** Sidebands generated by the pump beam at high powers when the system is pressurized with hydrogen at ~ 16.7 bar. $S_1$ and $AS_1$ denote the first Stokes and anti-Stokes sidebands of the pump. **d** Beams exiting the fiber when the transduction is turned on. With a hydrogen pressurization of ~ 16.7 bar, the process is phase-matched for up-conversion. Here, $M_1$ denotes the up-converted mixing sideband. **e** Sidebands of the pump present when the pressure is of ~ 28.1 bar. $S_2$ denotes the second Stokes band. **f** Image of the beams involved in the down-conversion process. $M_{-1}$ denotes the down-converted mixing band. **e** and **f** are photographed through an infrared viewer. **g** Same case as **f** but photographed in the visible spectrum as done for cases **a**-**d**.

Figs. 3a-d showcase the key role of the coherence waves in molecular modulation. At Fig. 3a, we can see the standalone profile of the mixing beam ($M_0$). In Fig. 3b, both the pump (P) and $M_0$ beams coexist in the same fiber, but they are both below the Raman threshold and are unable to initiate stimulated Raman scattering (SRS), as no sidebands can be observed. However, when the pump power is set above the threshold, it is able to initiate noise-seeded SRS and generate Raman sidebands as seen in Fig. 3c at a hydrogen pressure of ~ 16.7 bar and in Fig. 3e at ~ 28.1 bar, which indicates the presence of molecular coherence waves.

In this experiment, we specifically focused on molecular modulation by phase matching with the molecular coherence wave induced by the pump and Stokes ($S_1$) sidebands. In this regard, Fig. 3d showcases the up-conversion transduction process, in which the $M_0$ mixing (below threshold) experiences a frequency up-shift to the targeted $M_1$ sideband. Meanwhile, Figs. 3f-g correspond to the down-conversion process to the $M_{-1}$ sideband. All the photographies have been taken under the same camera settings, with exception of Figs. 3e-f being taken through an infrared viewer.

## 3 Theoretical modelling

In this section, we show the Maxwell-Bloch propagation simulations corresponding to the up-conversion (Fig. 4) and down-conversion (Fig. 5) processes. The simulations were done by considering Eqs. (2) to (4) of the main text with 5 sidebands for both pump and mixing pulses.

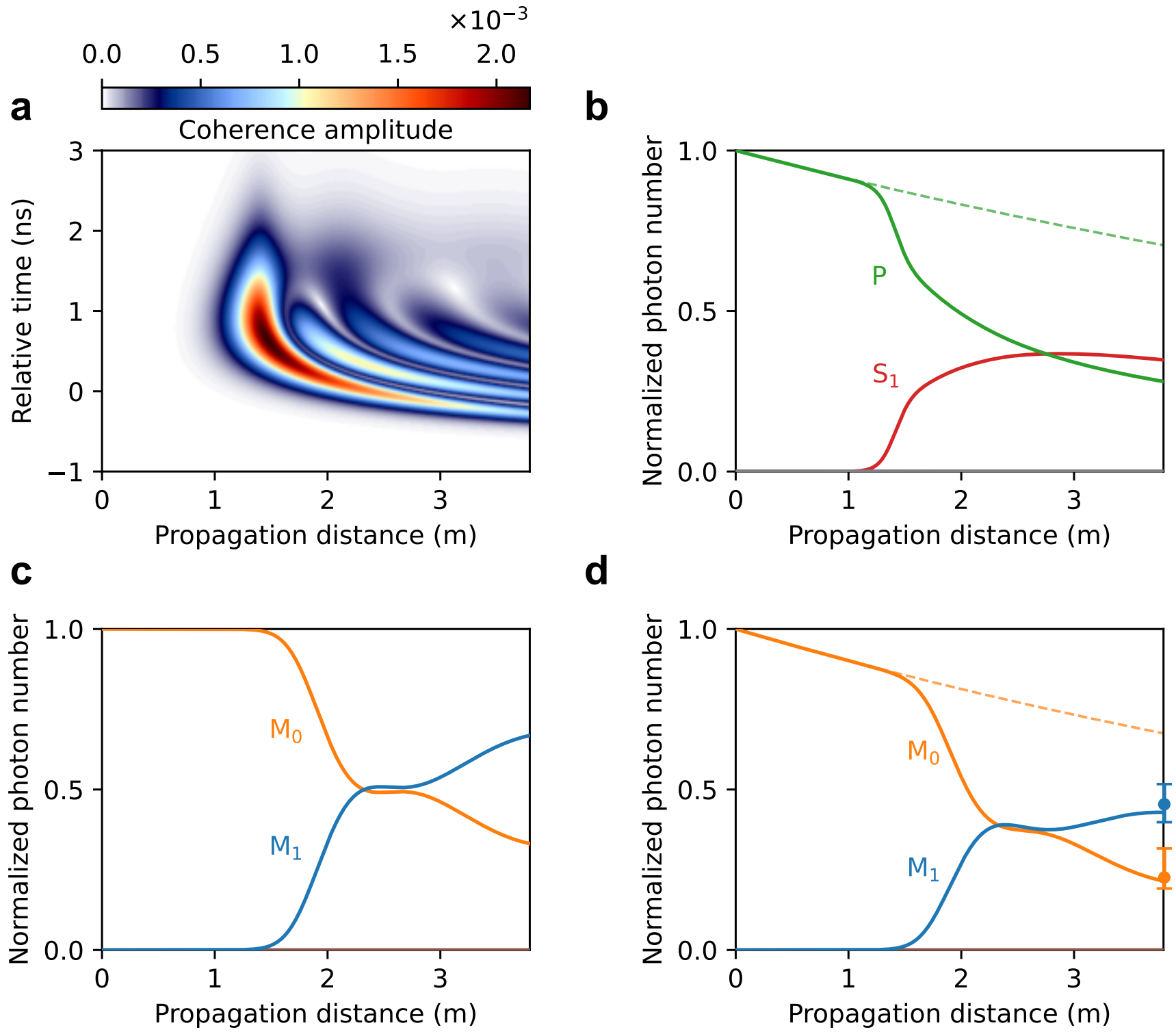

**Fig. 4** Maxwell-Bloch propagation simulations for up-conversion. **a** Evolution of the spatial distribution of the amplitude of the coherence wave induced by the pump and Stokes pulses. A slice in the vertical axis at a given propagation distance represents the distribution of the coherence wave in the time domain, measured relative to the propagating pulses. Note that negative times correspond to the leading edge of the pulses, while the trailing edge lies in the future (positive times). **b** Evolution of the average photon number population of the pump sidebands. The photon numbers are normalized with respect to the initial photon number population of the pump wavelength. Only the pump (P) and Stokes ($S_1$) bands are dominant, driving the molecular coherence wave at **a**. **c** and **d** show the evolution of the photon number population of the mixing sidebands, with $M_0$ and $M_1$ denoting the mixing and up-converted bands. In **c**, the propagation losses of the mixing sidebands are turned off, while **d** includes them. The dashed lines represent the evolution of the pump and mixing normalized photon number populations considering only guidance.

Figs. 4a and 5a show the evolution of the pulse envelope amplitude of the molecular coherence wave induced by the pump and Stokes frequencies, whose dynamics are shown at Figs. 4b and 5b. This amplitude would correspond to $|Q(z, \tau)|$, with $Q$ being the molecular coherence introduced at Eqs. (2) to (4) of the main text. Here, $z$ is the propagation distance and $\tau$ the relative time, which is the time in the reference frame co-propagating with the pulses. The pulses are considered to have the same group velocity (approximated to the speed of light in vacuum) due to the magnitude of their duration and the total distance traveled in the gaseous medium provided by the fiber.

The pump and mixing pulses are set initially to be synchronized gaussian envelopes of duration 1.42 ns and 0.3 ns respectively. Here, the 515 nm pump pulse energy is of $\sim 0.5$ µJ so that Figs. 4c and 5c correspond to the phase-matched transduction processes shown at the Fig. 4a in the main text. Meanwhile, the energy of the mixing pulse is set to $\sim 10$ nJ, well below the Raman threshold of the system estimated to be over 100 nJ. Note that the dynamics between the mixing sidebands will not change as long as the mixing input pulse energies are kept below the Raman threshold and the mixing is unable to generate its own molecular coherence. Therefore, at Figs. 4 and 5 we normalize the photon number axis to the input value.

At Figs. 4d and 5d, the experimental measurements for the population of each band are shown at the propagation distance mark of 3.7 m for comparison with the simulations. In this case, the dynamics are simulated also including the fiber losses from Fig. 2 into the mixing sidebands. The measured values are normalized with respect to the mixing population when the conversion is off, taking also the losses shown at Fig. 2 into consideration. These data points correspond to the efficiency measurements presented in Fig. 4b of the main text, with the errorbars covering the average values measured. To focus on the underlying physics of the molecular-modulation-based transducer, the internal quantum efficiency $\eta$ is measured relative to the $M_0$ population reaching the end of the transducer when the conversion is turned off, making the measurement independent of the losses of a particular setup.

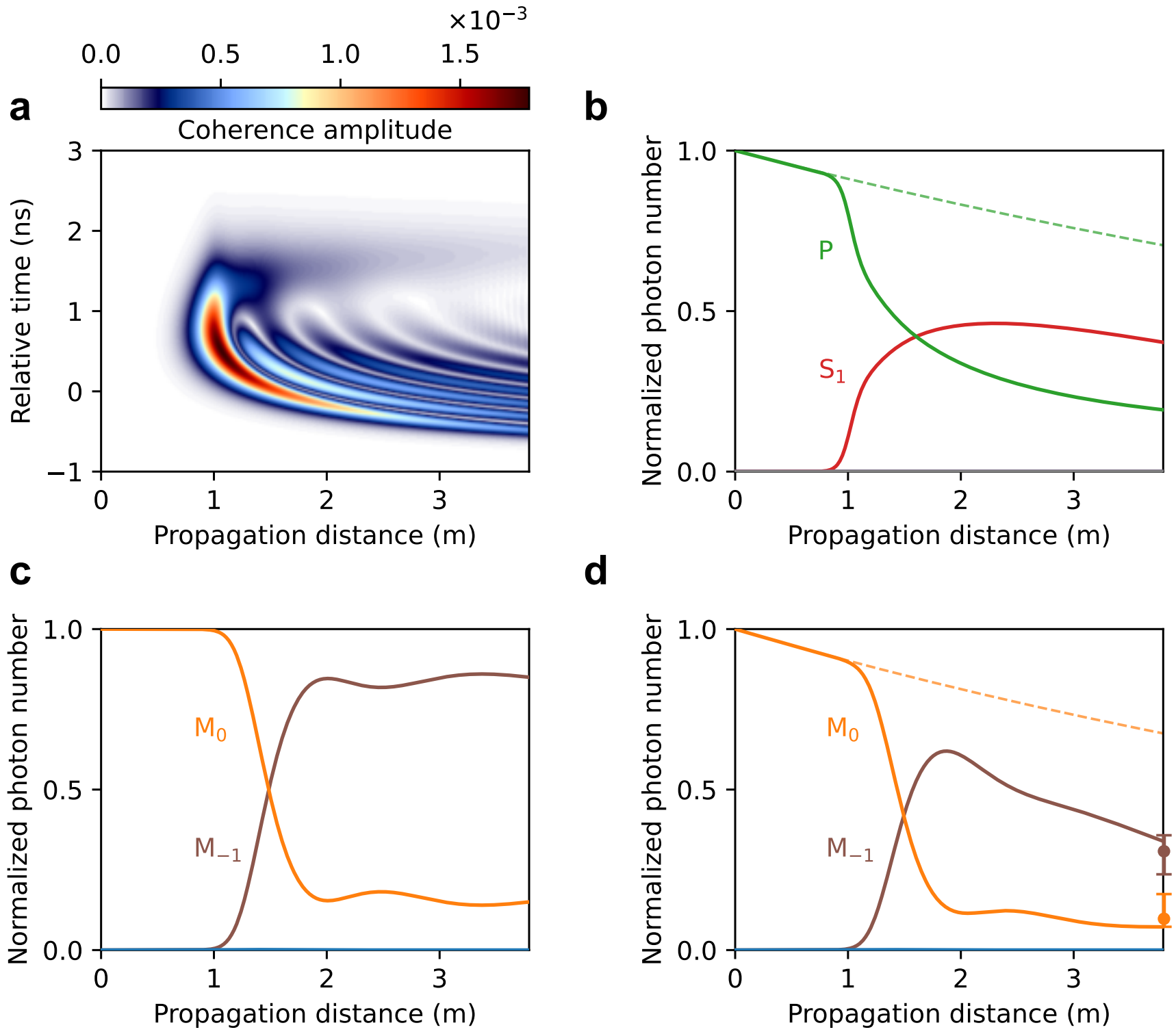

**Fig. 5** Maxwell-Bloch propagation simulations for down-conversion. This figure is analogous to Fig. 4. **a** Evolution of the spatial distribution of the amplitude of the coherence wave induced by the pump and Stokes pulses. **b** Evolution of the average photon number population of the pump sidebands. The photon numbers are normalized with respect to the initial photon number population of the pump wavelength. **c** and **d** show the evolution of the photon number population of the mixing sidebands, with $M_0$ and $M_{-1}$ denoting the mixing and down-converted bands. In **c**, the propagation losses of the mixing sidebands are turned off, while **d** includes them.

## 4 Unnormalized photon number distributions

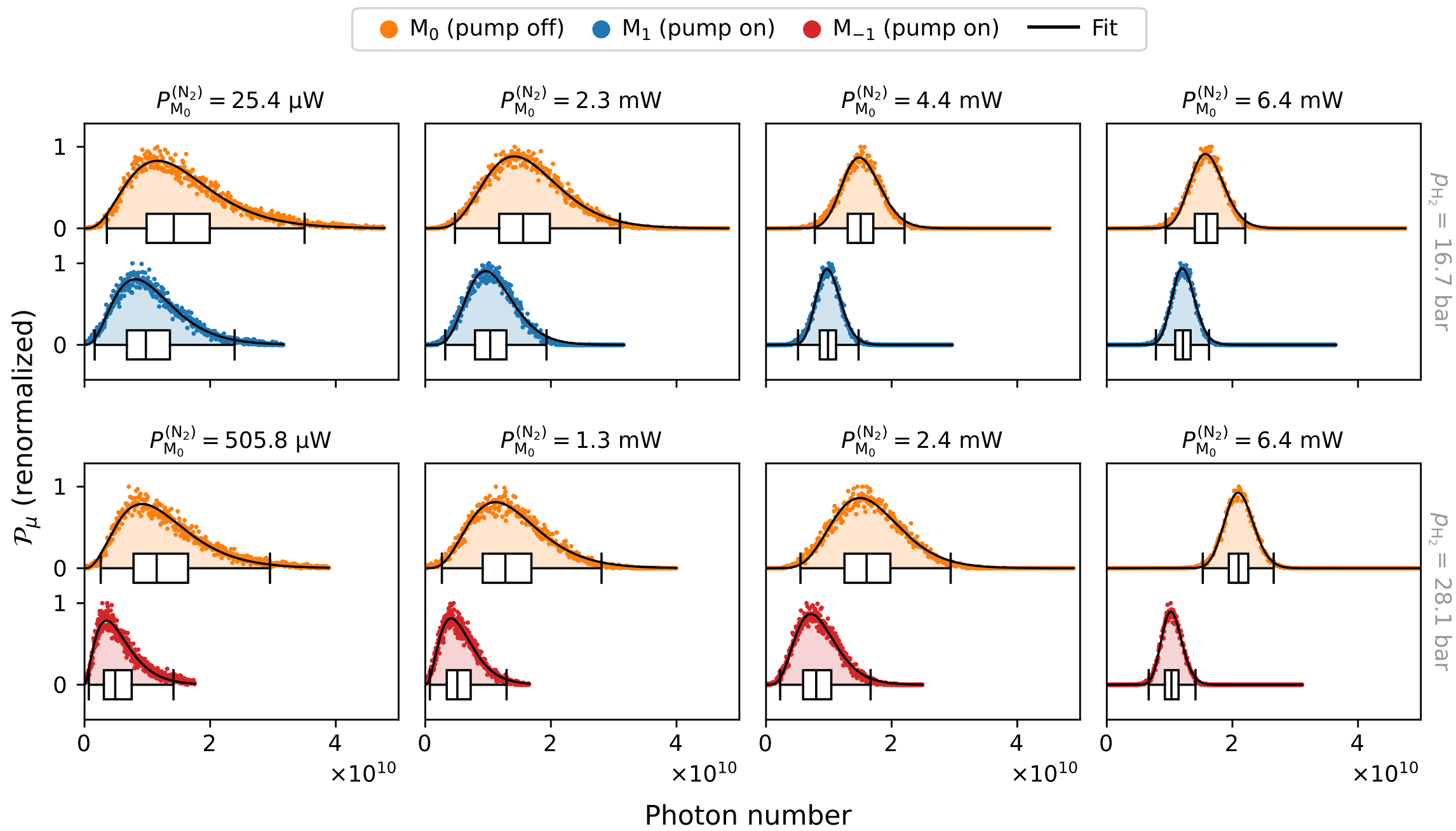

**Fig. 6** Unnormalized photon statistics for the conversion processes. The top row of subplots displays the up-conversion results, while the bottom row corresponds to down-conversion measurements. Each subplot shows the measurement of the mixing statistics when the transduction is turned off and the converted frequency statistics. The histograms are fitted to gamma distributions and the probabilities are renormalized to their maximum value. Box plots are shown below each histogram, indicating the quartiles and median of the distributions.

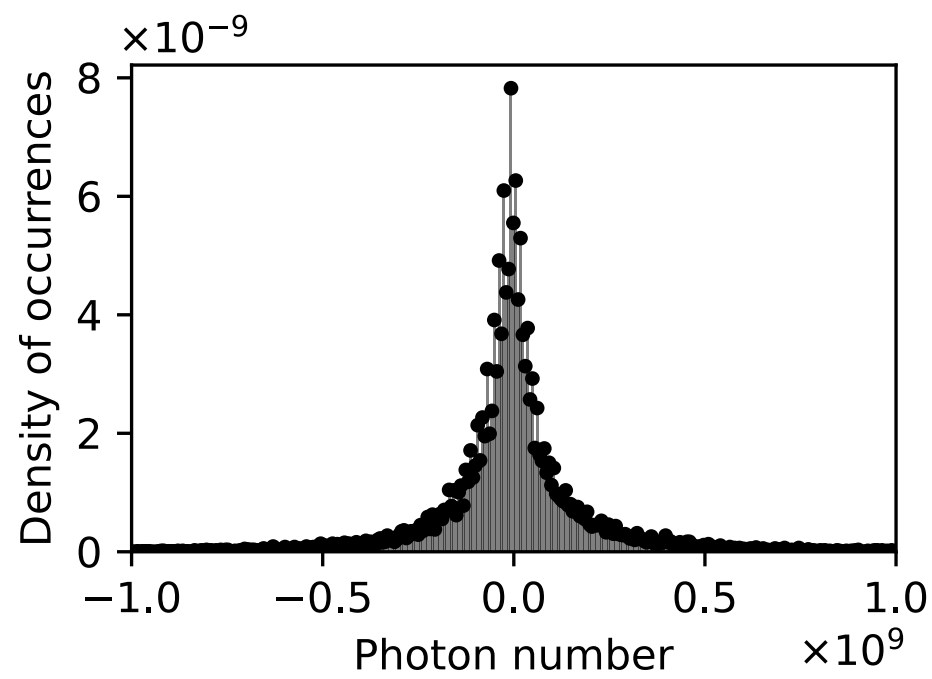

**Fig. 7** Example of a dark counts distribution. Here, the photon number axis corresponds to the 585 nm calibration.

## 5 Data processing

The histograms in the main text are fitted to gamma distributions of the form

$$\mathcal{P}(n;\mu,\langle n\rangle) = \frac{\mu^{\mu}}{\langle n\rangle\,\Gamma(\mu)}\left(\frac{n}{\langle n\rangle}\right)^{\mu-1}\exp\left(-\mu\frac{n}{\langle n\rangle}\right), \tag{1}$$

where $\mu$ is the number of statistically-independent modes involved in the detection and $\langle n\rangle$ is the average photon number.

### 5.1 Fit parameters and $g^{(2)}(0)$ functions

From the gamma distribution, with variance $\left\langle(\Delta n)^2\right\rangle = \langle n\rangle^2/\mu$, the zero-delay second order correlation function can be obtained as $g^{(2)}(0) = 1 + 1/\mu - 1/\langle n\rangle$. With this, Tables 1 and 2 show the resulting fit data for the up-conversion and down-conversion cases respectively. In these tables the uncertainties are obtained as specified in Section 6.

| | Mixing | | | Up-converted | | |
|---|---|---|---|---|---|---|
| $P_{\mathrm{M}_0}^{(\mathrm{N}_2)}$ | $\mu$ | $\langle n\rangle \times 10^{-10}$ | $g_{\mathrm{M}_0}^{(2)}(0)$ | $\mu$ | $\langle n\rangle \times 10^{-10}$ | $g_{\mathrm{M}_1}^{(2)}(0)$ |
| 25.4 µW | 4.17(6) | 1.531(7) | 1.23963243(3) | 3.99(5) | 1.080(5) | 1.25089010(3) |
| 2.3 mW | 6.80(8) | 1.659(5) | 1.147139501(15) | 7.83(11) | 1.097(3) | 1.127722379(18) |
| 4.4 mW | 25.7(4) | 1.544(2) | 1.038881507(5) | 26.7(3) | 1.0085(17) | 1.037438395(5) |
| 6.4 mW | 36.1(5) | 1.6145(17) | 1.027690297(3) | 45.2(5) | 1.230(1) | 1.022131646(2) |

**Table 1** Fit parameters and computed zero-delay second order correlation functions, $g^{(2)}(0)$, for all the up-conversion cases. Each case is labeled by $P_{\mathrm{M}_0}^{(\mathrm{N}_2)}$, which is the mixing ($\mathrm{M}_0$) average power at the output of the $N_2$-filled HCF before attenuation.

| | Mixing | | | Down-converted | | |
|---|---|---|---|---|---|---|
| $P_{\mathrm{M}_0}^{(\mathrm{N}_2)}$ | $\mu$ | $\langle n\rangle \times 10^{-10}$ | $g_{\mathrm{M}_0}^{(2)}(0)$ | $\mu$ | $\langle n\rangle \times 10^{-10}$ | $g_{\mathrm{M}_{-1}}^{(2)}(0)$ |
| 505.8 µW | 3.74(5) | 1.252(6) | 1.26763923(4) | 2.70(4) | 0.5552120(4) | 1.37046811(8) |
| 1.3 mW | 5.09(7) | 1.3891593(5) | 1.19664839(2) | 3.53(5) | 0.5744372(3) | 1.28368788(5) |
| 2.4 mW | 8.79(13) | 1.6864486(5) | 1.113728006(13) | 6.00(8) | 0.8569816(3) | 1.16674088(2) |
| 6.4 mW | 84.3(7) | 2.11519263(10) | 1.0118624512(7) | 40.9(4) | 1.044689(8) | 1.024455844(2) |

**Table 2** Fit parameters and computed zero-delay second order correlation functions, $g^{(2)}(0)$, for all the down-conversion cases. Each case is labeled by $P_{\mathrm{M}_0}^{(\mathrm{N}_2)}$, which is the mixing ($\mathrm{M}_0$) average power at the output of the $N_2$-filled HCF before attenuation.

## 5.2 Higher-order zero-delay correlation functions

Additionally, we can compute higher-order zero-delay correlation functions for the gamma distributions. The normally ordered third-order correlation function at zero delay is

$$g^{(3)}(0) = \frac{\left\langle (a^{\dagger})^3 a^3 \right\rangle}{\left\langle a^{\dagger} a \right\rangle^3}. \tag{2}$$

Using the ladder-operator commutation relation $[a, a^{\dagger}] = aa^{\dagger} - a^{\dagger}a = 1$, this can be expressed in terms of the photon-number moments $\langle n^r \rangle = \langle (a^{\dagger}a)^r \rangle$ as

$$g^{(3)}(0) = \frac{\left\langle n^3 \right\rangle - 3\left\langle n^2 \right\rangle + 2\left\langle n \right\rangle}{\left\langle n \right\rangle^3}. \tag{3}$$

For photon statistics following a gamma distribution like Equation (1), the raw moments are

$$\langle n^r \rangle = \left(\frac{\langle n \rangle}{\mu}\right)^r \frac{\Gamma(\mu + r)}{\Gamma(\mu)}. \tag{4}$$

Using the property $\Gamma(\mu + r) = \mu\Gamma(\mu + r - 1)$, they can be rewritten as

$$\langle n^r \rangle = \left(\frac{\langle n \rangle}{\mu}\right)^r \prod_{m=0}^{r-1} (\mu + m). \tag{5}$$

Substituting these moments into Equation (3) gives

$$g^{(3)}(0) = 1 + \frac{3}{\mu} + \frac{2}{\mu^2} - \frac{3}{\langle n \rangle} - \frac{3}{\mu \langle n \rangle} + \frac{2}{\langle n \rangle^2}. \tag{6}$$

For the zero-delay fourth-order correlation function, we start by considering the following formula:

$$g^{(4)}(0) = \frac{\left\langle (a^{\dagger})^4 a^4 \right\rangle}{\left\langle a^{\dagger} a \right\rangle^4}. \tag{7}$$

Applying the commutation properties of the ladder operators as previously done, in terms of the moments this formula becomes

$$g^{(4)}(0) = \frac{\left\langle n^4 \right\rangle - 6\left\langle n^3 \right\rangle + 11\left\langle n^2 \right\rangle - 6\left\langle n \right\rangle}{\left\langle n \right\rangle^4}. \tag{8}$$

For the gamma distribution in Equation (1), substitution of the raw moments yields

$$g^{(4)}(0) = 1 + \frac{6}{\mu} + \frac{11}{\mu^2} + \frac{6}{\mu^3} - \frac{6}{\langle n \rangle} - \frac{18}{\mu \langle n \rangle} - \frac{12}{\mu^2 \langle n \rangle} + \frac{11}{\langle n \rangle^2} + \frac{11}{\mu \langle n \rangle^2} - \frac{6}{\langle n \rangle^3}. \tag{9}$$

With the resulting Equations (6) and (9), we can compute the values of these functions using fit data and obtain the values presented at Tables 3 and 4.

| $P_{\mathrm{M}_0}^{(\mathrm{N}_2)}$ | $g_{\mathrm{M}_0}^{(3)}(0)$ | $g_{\mathrm{M}_1}^{(3)}(0)$ | $g_{\mathrm{M}_1}^{(3)}(0)/g_{\mathrm{M}_0}^{(3)}(0)$ | $g_{\mathrm{M}_0}^{(4)}(0)$ | $g_{\mathrm{M}_1}^{(4)}(0)$ | $g_{\mathrm{M}_1}^{(4)}(0)/g_{\mathrm{M}_0}^{(4)}(0)$ |
|---|---|---|---|---|---|---|
| 25.4 µW | 1.834(14) | 1.879(13) | 1.02(1) | 3.15(4) | 3.2(4) | 1.045(19) |
| 2.3 mW | 1.485(7) | 1.416(6) | 0.954(6) | 2.140(18) | 1.958(16) | 0.92(1) |
| 4.4 mW | 1.1197(18) | 1.1151(15) | 0.996(2) | 1.250(4) | 1.240(3) | 0.992(4) |
| 6.4 mW | 1.085(1) | 1.0674(7) | 0.9841(11) | 1.175(2) | 1.1382(15) | 0.969(2) |

**Table 3** Third and fourth-order zero-delay correlation functions retrieved from fit data in the case of frequency up-conversion.

| $P_{\mathrm{M}_0}^{(\mathrm{N}_2)}$ | $g_{\mathrm{M}_0}^{(3)}(0)$ | $g_{\mathrm{M}_{-1}}^{(3)}(0)$ | $g_{\mathrm{M}_{-1}}^{(3)}(0)/g_{\mathrm{M}_0}^{(3)}(0)$ | $g_{\mathrm{M}_0}^{(4)}(0)$ | $g_{\mathrm{M}_{-1}}^{(4)}(0)$ | $g_{\mathrm{M}_{-1}}^{(4)}(0)/g_{\mathrm{M}_0}^{(4)}(0)$ |
|---|---|---|---|---|---|---|
| 505.8 µW | 1.946(15) | 2.39(2) | 1.226(15) | 3.51(5) | 5.04(9) | 1.44(3) |
| 1.3 mW | 1.67(1) | 2.012(16) | 1.207(12) | 2.65(3) | 3.72(5) | 1.40(3) |
| 2.4 mW | 1.367(6) | 1.556(8) | 1.138(7) | 1.833(14) | 2.33(2) | 1.273(15) |
| 6.4 mW | 1.0359(3) | 1.0746(7) | 1.0374(8) | 1.0727(7) | 1.1534(15) | 1.0752(15) |

**Table 4** Third and fourth-order zero-delay correlation functions retrieved from fit data in the case of frequency down-conversion.

## 6 Propagation of uncertainty calculations

In this section we show how the uncertainties propagate from the photon number histogram fits to the metrics used to quantify the statistical similarity. For a given fit parameter $p$, we consider the standard deviation $\sigma_p$ as uncertainty.

### 6.1 $\mathcal{F}$ uncertainty

Let us consider the fidelity

$$.\mathcal{F} = \left(\mathrm{tr}\sqrt{\sqrt{\rho}\sigma\sqrt{\rho}}\right)^2, \tag{10}$$

where $\rho$ and $\sigma$ are the density matrices of the states that we are comparing. Specifically, we consider them in the normalized photon number space. Assuming that we are dealing with thermal states with diagonal density matrix, the diagonal elements of these matrices can be obtained directly from the probability density histograms. Therefore, considering two states

$$\rho = \sum_n \mathcal{P}(n; \mu_1, \langle n \rangle_1) \left|n\middle\rangle\middle\langle n\right|, \tag{11}$$

$$\sigma = \sum_n \mathcal{P}(n; \mu_2, \langle n \rangle_2) \left|n\middle\rangle\middle\langle n\right| \tag{12}$$

with respective average photon numbers $\langle n \rangle_1$ and $\langle n \rangle_2$, we write the fidelity as

$$.\mathcal{F}(\mu_1, \langle n \rangle_1, \mu_2, \langle n \rangle_2) = \left(\sum_n \sqrt{\mathcal{P}(n; \mu_1, \langle n \rangle_1)\mathcal{P}(n; \mu_2, \langle n \rangle_2)}\right)^2. \tag{13}$$

With a formula for fidelity depending on four parameters, its variance is given by

$$\begin{aligned}\sigma_{\mathcal{F}}^2 = &\left(\frac{\partial \mathcal{F}}{\partial \mu_1}\right)^2 \sigma_{\mu_1}^2 + \left(\frac{\partial \mathcal{F}}{\partial \langle n \rangle_1}\right)^2 \sigma_{\langle n \rangle_1}^2 + 2\frac{\partial \mathcal{F}}{\partial \mu_1}\frac{\partial \mathcal{F}}{\partial \langle n \rangle_1} \mathrm{Cov}(\mu_1, \langle n \rangle_1) \\ &+ \left(\frac{\partial \mathcal{F}}{\partial \mu_2}\right)^2 \sigma_{\mu_2}^2 + \left(\frac{\partial \mathcal{F}}{\partial \langle n \rangle_2}\right)^2 \sigma_{\langle n \rangle_2}^2 + 2\frac{\partial \mathcal{F}}{\partial \mu_2}\frac{\partial \mathcal{F}}{\partial \langle n \rangle_2} \mathrm{Cov}(\mu_2, \langle n \rangle_2),\end{aligned} \tag{14}$$

with Cov denoting the covariance between two parameters. This formula can be compacted with matrix formulation. With two fit parameters per distribution, we write the vectors

$$\theta_{\mathbf{m}} = \begin{pmatrix} \mu_m \\ \langle n \rangle_m \end{pmatrix}, \tag{15}$$

where $m = 1, 2$. Then, the covariance matrices from the fits are written as

$$C_m = \begin{pmatrix} \sigma_{\mu_m}^2 & \mathrm{Cov}(\mu_m, \langle n \rangle_m) \\ \mathrm{Cov}(\mu_m, \langle n \rangle_m) & \sigma_{\langle n \rangle_m}^2 \end{pmatrix}. \tag{16}$$

The elements of the $C_m$ matrices are obtained by fitting the photon-number histograms to Equation (1) using the `curve_fit` function from the SciPy library for the Python programming language.

If we now introduce the Jacobian

$$\nabla\mathcal{F} = \begin{pmatrix} \dfrac{\partial\mathcal{F}}{\partial\mu_1} \\ \dfrac{\partial\mathcal{F}}{\partial\langle n\rangle_1} \\ \dfrac{\partial\mathcal{F}}{\partial\mu_2} \\ \dfrac{\partial\mathcal{F}}{\partial\langle n\rangle_2} \end{pmatrix} \tag{17}$$

and, considering that the fits are statistically independent, the matrix

$$C = \begin{pmatrix} C_1 & 0 \\ 0 & C_2 \end{pmatrix}, \tag{18}$$

the standard deviation of the fidelity becomes

$$\sigma_{\mathcal{F}} = \sqrt{\nabla\mathcal{F}^{\mathrm{T}} C \nabla\mathcal{F}}. \tag{19}$$

Meanwhile, the partial derivatives in Equation (17) are computed by central finite differences:

$$\frac{\partial\mathcal{F}}{\partial\phi_i} = \frac{\mathcal{F}(\phi_i + h) - \mathcal{F}(\phi_i - h)}{2h}, \tag{20}$$

where $\phi_i$ denotes any of the four parameters in which the fidelity depends on and $h$ is a small step taken in the space of the individual parameter being varied to compute the derivative while the rest of the parameters are kept fixed. To obtain the numerator in this formula, the gamma distribution fit lines are recomputed with this parametric variation and then the corresponding fidelity values are calculated.

## 6.2 $g^{(2)}(0)$ uncertainty

Considering that the distribution at Equation (1) has variance $\left\langle (\Delta n)^2 \right\rangle = \langle n\rangle^2/\mu$, the values of the zero-delay second order correlation function can be obtained as

$$g^{(2)}(0) = 1 + \frac{1}{\mu} - \frac{1}{\langle n\rangle}. \tag{21}$$

With this, the variance is given by

$$\sigma^2_{g^{(2)}(0)} = \left(\frac{\partial g^{(2)}(0)}{\partial\mu}\right)^2 \sigma^2_\mu + \left(\frac{\partial g^{(2)}(0)}{\partial\langle n\rangle}\right)^2 \sigma^2_{\langle n\rangle} + 2\frac{\partial g^{(2)}(0)}{\partial\mu}\frac{\partial g^{(2)}(0)}{\partial\langle n\rangle}\,\mathrm{Cov}(\mu,\langle n\rangle), \tag{22}$$

which evaluates as

$$\sigma^2_{g^{(2)}(0)} = \frac{\sigma^2_\mu}{\mu^4} + \frac{\sigma^2_{\langle n\rangle}}{\langle n\rangle^4} - \frac{2\,\mathrm{Cov}(\mu,\langle n\rangle)}{\mu^2\langle n\rangle^2}. \tag{23}$$

## 6.3 $g^{(3)}(0)$ uncertainty

Considering the form of the zero-delay third-order correlation function at Equation (6), the propagation formula for the uncertainty $\sigma_{g^{(3)}(0)}$ is given by

$$\sigma^2_{g^{(3)}(0)} = \left(\frac{\partial g^{(3)}(0)}{\partial\langle n\rangle}\right)^2 \sigma^2_{\langle n\rangle} + \left(\frac{\partial g^{(3)}(0)}{\partial\mu}\right)^2 \sigma^2_\mu + 2\frac{\partial g^{(3)}(0)}{\partial\langle n\rangle}\frac{\partial g^{(3)}(0)}{\partial\mu}\,\mathrm{Cov}(\mu,\langle n\rangle). \tag{24}$$

Therefore, the required derivatives are

$$\frac{\partial g^{(3)}(0)}{\partial \langle n \rangle} = \frac{3}{\langle n \rangle^2}\left(1+\frac{1}{\mu}\right) - \frac{4}{\langle n \rangle^3}, \tag{25}$$

$$\frac{\partial g^{(3)}(0)}{\partial \mu} = -\frac{3}{\mu^2} - \frac{4}{\mu^3} + \frac{3}{\mu^2 \langle n \rangle}. \tag{26}$$

With these, we write

$$\sigma^2_{g^{(3)}(0)} = \left[\frac{3}{\langle n \rangle^2}\left(1+\frac{1}{\mu}\right) - \frac{4}{\langle n \rangle^3}\right]^2 \sigma^2_{\langle n \rangle} + \left[-\frac{3}{\mu^2} - \frac{4}{\mu^3} + \frac{3}{\mu^2 \langle n \rangle}\right]^2 \sigma^2_{\mu} + 2\left[\frac{3}{\langle n \rangle^2}\left(1+\frac{1}{\mu}\right) - \frac{4}{\langle n \rangle^3}\right]\left[-\frac{3}{\mu^2} - \frac{4}{\mu^3} + \frac{3}{\mu^2 \langle n \rangle}\right] \mathrm{Cov}(\mu, \langle n \rangle). \tag{27}$$

## 6.4 $g^{(4)}(0)$ uncertainty

The propagated variance of $g^{(4)}$ is

$$\sigma^2_{g^{(4)}(0)} = \left(\frac{\partial g^{(4)}(0)}{\partial \langle n \rangle}\right)^2 \sigma^2_{\langle n \rangle} + \left(\frac{\partial g^{(4)}(0)}{\partial \mu}\right)^2 \sigma^2_{\mu} + 2\frac{\partial g^{(4)}(0)}{\partial \langle n \rangle}\frac{\partial g^{(4)}(0)}{\partial \mu} \mathrm{Cov}(\mu, \langle n \rangle), \tag{28}$$

where the required partial derivatives, computed considering Equation (9), are

$$\frac{\partial g^{(4)}(0)}{\partial \langle n \rangle} = \frac{6}{\langle n \rangle^2} + \frac{18}{\mu \langle n \rangle^2} + \frac{12}{\mu^2 \langle n \rangle^2} - \frac{22}{\langle n \rangle^3} - \frac{22}{\mu \langle n \rangle^3} + \frac{18}{\langle n \rangle^4}, \tag{29}$$

$$\frac{\partial g^{(4)}(0)}{\partial \mu} = -\frac{6}{\mu^2} - \frac{22}{\mu^3} - \frac{18}{\mu^4} + \frac{18}{\mu^2 \langle n \rangle} + \frac{24}{\mu^3 \langle n \rangle} - \frac{11}{\mu^2 \langle n \rangle^2}. \tag{30}$$

## 6.5 $g^{(n)}_{\mathrm{M}_{\pm 1}}(0)/g^{(n)}_{\mathrm{M}_0}(0)$ uncertainty

To compute the uncertainty in the ratio of zero-delay correlation functions $R^{(n)}_{\pm} = g^{(n)}_{\mathrm{M}_{\pm 1}}(0)/g^{(n)}_{\mathrm{M}_0}(0)$, with $n = 2, 3, 4$, we can directly use the variances previously computed considering that the covariance between independent datasets is zero:

$$\sigma^2_{R^{(n)}_{\pm}} = \left(\frac{\partial R^{(n)}_{\pm}}{\partial g^{(n)}_{\mathrm{M}_{\pm 1}}(0)}\right)^2 \sigma^2_{g^{(n)}_{\mathrm{M}_{\pm 1}}(0)} + \left(\frac{\partial R^{(n)}_{\pm}}{\partial g^{(n)}_{\mathrm{M}_0}(0)}\right)^2 \sigma^2_{g^{(n)}_{\mathrm{M}_0}(0)} = \frac{1}{[g^{(n)}_{\mathrm{M}_0}(0)]^2}\sigma^2_{g^{(n)}_{\mathrm{M}_{\pm 1}}(0)} + \frac{[g^{(n)}_{\mathrm{M}_{\pm 1}}(0)]^2}{[g^{(n)}_{\mathrm{M}_0}(0)]^4}\sigma^2_{g^{(n)}_{\mathrm{M}_0}(0)}. \tag{31}$$